\documentclass[12pt]{article}

\usepackage{latexsym}
\usepackage{amssymb,amsfonts,amsmath}
\usepackage{graphicx} 
\usepackage{indentfirst}
\usepackage{bbm}
\usepackage{amssymb}
\usepackage{verbatim}
\usepackage{amsmath, amsthm,amssymb}
\usepackage{mathrsfs}
\usepackage{hyperref}
\usepackage{amsfonts}
\usepackage{dsfont}
\usepackage{cite}
\usepackage{tikz}
\usepackage{enumitem}

\parskip=\medskipamount

\newcommand {\cA}{{\cal A}}
\newcommand {\cB}{{\cal B}}

\newcommand {\cD}{{\cal D}}

\newcommand {\cI}{{\cal I}}

\newcommand {\cK}{{\cal K}}

\newcommand {\cN}{{\cal N}}
\newcommand {\cO}{{\cal O}}

\newcommand {\cQ}{{\cal Q}}

\newcommand {\cV}{{\cal V}}

\newcommand{\bX}{{\bf X}}

\def\a{\alpha}

\def\b{\beta}
\def\c{\chi}
\def\d{\delta}

\def\g{\gamma}

\def\l{\lambda}
\def\m{\mu}

\def\o{\omega}

\def\q{\theta}
\def\r{\rho}
\def\s{\sigma}

\def\x{\xi}

\def\F{\Phi}
\def\J{\Psi}

\def\P{\Pi}
\def\Q{\Theta}

\def\ri{{\rm i}}

\def\ra{{\rm a}}

\newcommand{\ad}{{\dot{\alpha}}}                           
\newcommand{\bd}{{\dot{\beta}}}                            
\newcommand{\ve}{\varepsilon}                            

\newcommand{\pa}{\partial}                           
\newcommand{\hf}{\frac12}

\newcommand{\be}{\begin{equation}}
\newcommand{\ee}{\end{equation}}
\newcommand{\bea}{\begin{eqnarray}}
\newcommand{\eea}{\end{eqnarray}}
\newcommand{\non}{\nonumber}

\newcommand{\bm}[1]{\mbox{\boldmath$#1$}}

\def\double #1{#1{\hbox{\kern-2pt $#1$}}}

\newcommand{\gd}{{\dot\g}}

\newif\ifdtup

\def\la{{\langle}}
\def\ra{{\rangle}}
\def\fn3{{\bm {X}_{3}, \bm {\Q}_{3}, \bar {\bm \Q}_{3}}}
\def\fxq{{\bm {X}, \bm {\Q}, \bar {\bm \Q}}}
\def\corr1{{\la \bar S_{\ad(k)}(z_1) S_{\a(k)}(z_2) L(z_3) \ra}}
\def\cor2{{\la \bar S_{\ad(k)}(z_1) S_{\a(k)}(z_2) J_{\g \gd}(z_3) \ra}}
\def\c3{{\la \bar S_{\ad(k)}(z_1) S_{\b(k+l)}(z_2) \bar{S}_{\gd(l)}(z_3) \ra}}

\newcommand{\bsubeq}{\begin{subequations}}
\newcommand{\esubeq}{\end{subequations}}

\numberwithin{equation}{section}

\newcommand{\sU}{\mathsf{U}}

\begin{document}

\begin{titlepage}
\begin{flushright}
June, 2021 \\
\end{flushright}
\vspace{5mm}

\begin{center}
{\Large \bf 
Three-point functions of higher-spin spinor current multiplets in ${\mathcal N}=1$ superconformal theory}
\\ 
\end{center}

\begin{center}

{\bf
Evgeny I. Buchbinder, Jessica Hutomo
and Sergei M. Kuzenko} \\
\vspace{5mm}

\footnotesize{
{\it Department of Physics M013, The University of Western Australia\\
35 Stirling Highway, Crawley W.A. 6009, Australia}}  
~\\

\vspace{2mm}
~\\
\texttt{Email: evgeny.buchbinder@uwa.edu.au,
jessica.hutomo@uwa.edu.au, sergei.kuzenko@uwa.edu.au}
\vspace{2mm}

\end{center}

\begin{abstract}
\baselineskip=14pt

In this paper, we study the general form of three-point functions of conserved current multiplets $S_{\alpha(k)}= S_{(\alpha_1 \dots \alpha_k)}$ 
of arbitrary rank in four-dimensional 
${\mathcal N}=1$ superconformal theory. We find that the correlation function of three such operators 
$\langle \bar{S}_{\dot{\alpha}(k)} (z_1) S_{\beta(k+l)} (z_2) \bar{S}_{\dot{\gamma}(l)} (z_3) \rangle$ 
is fixed by the superconformal symmetry up to a single complex coefficient though the precise form of the correlator 
depends on the 
values of
$k$ and $l$.
In addition, we present the general structure of mixed correlators of the form 
$\langle \bar{S}_{\dot{\alpha}(k)} (z_1) S_{\alpha(k)} (z_2) L(z_3) \rangle$ 
and 
$\langle \bar{S}_{\dot{\alpha}(k)} (z_1) S_{\alpha(k)} (z_2) J_{\gamma \dot{\gamma}} (z_3) \rangle$, where $L$ is the flavour current multiplet and $J_{\gamma \dot{\gamma}}$ is the supercurrent.

\end{abstract}
\vspace{5mm}

\vfill
\end{titlepage}

\newpage
\renewcommand{\thefootnote}{\arabic{footnote}}
\setcounter{footnote}{0}

\tableofcontents{}
\vspace{1cm}
\bigskip\hrule

\allowdisplaybreaks


\section{Introduction}


It is well known that in (super)conformal field theory the general form of two- and 
three-point functions of conserved currents is fully determined by 
the (super)conformal symmetry and conservation laws up to finitely many independent coefficients. 
In the non-supersymmetric case, the systematic study of the two- and three-point functions of conserved currents was presented in~\cite{OP, EO} 
building on the earlier results obtained in Refs.~\cite{Polyakov:1970xd, Schreier:1971um, Migdal:1971xh, Migdal:1972tk, Ferrara:1972cq, Ferrara:1973yt, Koller:1974ut, Mack:1976pa, Stanev:1988ft, Fradkin:1978pp}.\footnote{Parity violating structures
in three-dimensional conformal field theory were not considered  in~\cite{OP, EO}. 
They  were later found in~\cite{Giombi:2011rz}.} 
The approach of~\cite{OP, EO} was later extended to superconformal field theories in 
diverse dimensions in~\cite{Park1, OsbornN1, Park, Park6, Park3, KT, Nizami:2013tpa, Buchbinder:2015qsa, Buchbinder:2015wia, 
Kuzenko:2016cmf, Buchbinder:2021gwu}.

The most important conserved currents in conformal field theory are the energy-momentum tensor and vector currents. 
In the supersymmetric case, they are embedded in special multiplet of conserved currents.
The energy-momentum tensor is, therefore, replaced with the supercurrent~\cite{Ferrara:1974pz} (see also \cite{GGRS, MSW, KS2, K-var, K-var1}) 
and vector currents are replaced with flavour current multiplets~\cite{FWZ}.
Three-point functions of these current multiplets have been extensively studied. In general, (super)conformal field theories also possess higher-spin conserved currents. 
In the case of three-dimensional conformal field theory, it was proven (under certain assumptions) 
by Maldacena and Zhiboedov in~\cite{Maldacena:2011jn} that
all correlation functions of higher-spin currents are equal to the ones in a free theory.
The theorem of Maldacena and Zhiboedov was later generalised by Stanev~\cite{Stanev:2013qra} and by  
Alba and Diab~\cite{Alba:2013yda, Alba:2015upa} to the four- (and higher-) dimensional case. 
One can view these results as the analog of the Coleman-Mandula theorem~\cite{Coleman:1967ad} for conformal field theories.

We believe that the analysis of~\cite{Maldacena:2011jn, Stanev:2013qra, Alba:2013yda, Alba:2015upa} has some limitations. 
First, the authors of~\cite{Maldacena:2011jn, Stanev:2013qra, Alba:2013yda, Alba:2015upa}
considered only bosonic symmetric traceless currents.
However, in supersymmetric theories conserved currents form supermultiplets consisting of both bosonic 
and fermionic component currents. Second, the results of~\cite{Maldacena:2011jn, Stanev:2013qra, Alba:2013yda, Alba:2015upa} are proven under certain assumptions, 
the main being the existence of only one conserved current of spin two which is the energy-momentum tensor. In~\cite{Maldacena:2011jn} 
it was shown that in three-dimensional conformal field theory the existence of a half-integer higher-spin conserved current implies the existence 
of one more conserved current of spin two. This means that in supersymmetric conformal field theory possessing higher-spin currents the
assumptions of~\cite{Maldacena:2011jn} might be violated. It is likely that the same conclusion also holds in four dimensions. 
Therefore, in the supersymmetric case it is unclear if higher-spin current multiplets exist only in free theory. 
In any case, regardless whether a theory is free or not,  it is interesting to understand the general structure of their correlation functions. 

In the non-supersymmetric case, the general structure of the three-point functions of conserved bosonic, vector currents of arbitrary spin was determined by 
Stanev \cite{Stanev:2012nq} and  Zhiboedov \cite{Zhiboedov:2012bm}, 
see also \cite{EKS} for similar results in the embedding formalism.
The aim of this paper is to make first steps towards determining the three-point functions of 
conserved higher-spin currents 
in the $\cN=1$ supersymmetric case in four dimensions. In fact, we will be interested in multiplets which do not contain vector currents.
To start with, it is worth reminding the reader of the known conformal current multiplets. 

In the case of $\cN=1$ supersymmetry, there are three types of conformal current multiplets depending on the corresponding superfield Lorentz type $(k/2,l/2)$. 
These multiplets were explicitly described in flat \cite{Ceresole:1999zs} and curved \cite{KR} backgrounds.
\begin{itemize} 
\item Given two positive integers $k$ and $l $, the 
conformal current multiplet $J_{\a(k) \ad(l)}:= J_{\a_1 \dots \a_k\ad_1 \dots \ad_l}
=J_{(\a_1 \dots \a_k) (\ad_1 \dots \ad_l)}$ obeys the conservation equations
\bea
D^\b J_{\a(k-1) \b \ad(l)} = 0~, \qquad \bar{D}^\bd J_{\a(k) \ad(l-1) \bd} = 0 ~.
\eea
This superfield has dimension $\big(2+ \hf (k+l) \big)$ and its $\sU(1)_R$ charge is 
equal to $\frac 13 (k -l) $, see \cite{KR} for the technical details. 
If $k=l$, the supercurrent $ J_{\a(k) \ad(k)} $ is neutral with respect to 
the $R$-symmetry group $\sU(1)_{R}$, and therefore it is consistent to 
restrict  $ J_{\a(k) \ad(k)} $ to be real. The $k=l=1$ case corresponds to the ordinary 
conformal supercurrent \cite{Ferrara:1974pz}. The case $k=l>1$ was first discussed 
in \cite{HST81}.

\item If $k>0$ and $l=0$, the conformal current multiplet $S_{\a(k)}$ obeys the 
the conservation equation
\bea
D^{\b}S_{\b\a_1 \dots \a_{k-1}} = 0\,,  \qquad  \bar D^2 S_{\a(k)} = 0~.
\label{zh02}
\eea
The case $k=1$ was first considered in \cite{KT}, where it was shown that the spinor supercurrent 
$S_\a$ naturally originates from the reduction of the conformal $\cN=2$ supercurrent \cite{Sohnius79}
to $\cN=1$ superspace. 

\item Finally, the $k=l=0$ case corresponds to the flavour current multiplet \cite{FWZ}, 
$L=\bar L$, constrained by 
\bea
		D^2 L = 0~, \qquad \bar{D}^2 L = 0 ~.
\eea
\end{itemize}
Do all of these multiplets occur in superconformal field theories?

The conformal supercurrent $J_{\a\ad}$ exists in every $\cN=1$ superconformal field theory. Flavour current multiplets exists in every superconformal field theory possessing an internal symmetry Lie group. The spinor current multiplet $S_\a$ exists in every $\cN=2$ superconformal field theory realised in terms of $\cN=1$ superfields. 
Explicit realisations in terms of free conformal scalar multiplets are known for the conformal current higher-spin multiplets $J_{\a(k) \ad(k)}$ 
and $J_{\a(k+1) \ad(k)}$, with $k>1$, both in Minkowski  \cite{KMT} and anti-de Sitter \cite{BHK18} superspace.

Higher-spin current multiplets $J_{\a(k) \ad(l)}$ and $S_{\a(k)} $ may be realised in terms of the on-shell chiral field strengths $W_{\a(k)} $ and 
massless antichiral scalar $\bar \F$ constrained as
\begin{subequations}
\bea
\bar D_\bd W_{\a(k)} &=&0~, \qquad D^\b W_{\b \a(k-1)} =0~, \label{1.4a} \\
D_\b \bar \F &=&0 ~, \qquad \bar D^2 \bar \F=0~.
\eea
\end{subequations}
These realisations are as follows \cite{BHK18,BGK18, GK19}:
\bea
J_{\a(k) \ad(l) } =W_{\a(k)} \bar W_{\ad(l)} ~, \qquad 
S_{\a(k)} =W_{\a(k)} \bar \F~.
\label{1.5}
\eea
Here $W_\a$, $W_{\a(2)}$ and $W_{\a(3)}$ are the gauge-invariant field strength describing the on-shell vector, gravitino and linearised supergravity multiplets, respectively.  
The chiral superfields $W_{\a (k)}$ for $k>3$ are the on-shell gauge-invariant field strengths corresponding to the massless higher-spin gauge multiplets 
\cite{KPS,KS93} (see section 6.9 in \cite{Ideas} for a review). Choosing $k=l=1$ in 
\eqref{1.5} gives the supercurrent of the free $\cN=1$ vector multiplet, 
$J_{\a \ad } =W_{\a} \bar W_{\ad}$, and the spinor supercurrent of the free $\cN=2$ vector multiplet, $S_{\a} =W_{\a} \bar \F$.

The off-shell gauge-invariant models for massless higher-spin multiplets proposed in 
\cite{KPS,KS93} are not superconformal, although the corresponding on-shell field strengths $W_{\a(k)}$ constrained by \eqref{1.4a} furnish irreducible representations of the superconformal group, see \cite{Ideas} for the technical details. At the moment we are not aware of explicit realisations of the current multiplets 
$S_{\a(k)} $ for $k>1$ in superconformal field theories. Nevertheless, it is of  interest to understand
the most general structure of three-point correlation functions of such current multiplets, which are compatible with the superconformal symmetry and conservation laws.

In this paper we will restrict ourselves to four-dimensional ${\cal N}=1$ superconformal field theory and to higher-spin 
current multiplets carrying only undotted or dotted indices, 
$S_{\a(k)} $ and its conjugate ${\bar S_{\ad(k)}} $.
We will refer to them as ``higher-spin spinor current multiplets". For $k=1$ the multiplet $S_{\a}$ is just a spinor current multiplet. 
This case was studied in our previous work~\cite{BHK21}. Three-point correlation functions
of more general higher-spin current multiplets require additional study and will be explored elsewhere. 

The main result of this paper is the most general structure of the three-point function 
\be 
\c3
\label{zh03}
\ee
for arbitrary integers $k$ and $\ell$. We showed that in all cases eq.~\eqref{zh03} is fixed by the superconformal symmetry and the 
conservation equations~\eqref{zh02} up to a single overall 
coefficient. However, we found that one has to distinguish two cases depending on values of $k$ and $\ell$, see section \ref{Section5} for details. 
Additionally, we also studied the mixed correlators
\be
\corr1\,, \qquad \cor2~, 
\label{zh04}
\ee
where $L$ is the flavour current multiplet and $J_{\g \gd}$ is the supercurrent. 

The paper is organised as follows. A brief review of the two- and three-point building blocks for correlation functions is given in section \ref{Section2}. 
In section \ref{Section3} we study a three-point 
function of higher-spin spinor current multiplets with the flavour current multiplet. We show that this three-point function is fixed by the superconformal 
symmetry up to two independent real coefficients. In the special case $k=1$, our result coincides with that found in our previous work~\cite{BHK21}. 
In section \ref{Section4} we consider a three-point function of higher-spin spinor current multiplets with the supercurrent. 
In this case the general form of the correlation function is fixed up to three independent real coefficients. In the special case $k=1$, our result coincides with that found in~\cite{BHK21}. Finally, 
in section \ref{Section5} we compute three-point correlator involving just the conserved higher-spin spinor current multiplets with arbitrary number of 
dotted and undotted indices. We found that its general structure is fixed up to a single complex coefficient.


\section{Superconformal building blocks} \label{Section2}


This section contains a concise summary of two and three-point superconformal building blocks in 4D $\cN=1$ superspace, which are important for our subsequent analysis. These superconformal structures were introduced in \cite{Park1, OsbornN1}, and later generalised to arbitrary ${\cN}$ in \cite{Park} (see also \cite{KT} for a review). 
A review of the general structure of two- and three-point correlation functions of primary operators is also given.
The presentation of this section closely follows \cite{BHK21}. Our 4D notation and conventions are those of \cite{Ideas}.


\subsection{Infinitesimal superconformal transformations}


We denote the $\cN=1$ Minkowski superspace by ${\mathbb M}^{4|4}$. It is parametrised by the local coordinates $z^A = (x^a, \q^\a, {\bar \q}_\ad)$, where $a = 0,1, \cdots 3,~ \a, \ad = 1, 2$. 
Infinitesimal superconformal transformations 
\be
\d z^A = \xi z^A~,
\ee
are generated by conformal Killing real supervector fields \cite{Ideas, KT}
\be
\x = {\overline \x} = \x^a (z) \pa_a + \x^\a (z)D_\a
+ {\bar \x}_\ad (z) {\bar D}^\ad
\ee   
obeying the equation
\be
[\x \, , \, D_\a ] ~\propto ~ D_\b ~.
\label{Keqn}
\ee   
As a result, the spinor parameters are expressed in terms of the vector ones
\be
\x^\a = -\frac{\rm i}{8} {\bar D}_{\bd} \x^{\bd \a}\;, \qquad
{\bar D}_{\bd} \x^\a = 0~.
\label{spinsc}
\ee
The latter satisfy
\be
D_{(\a} \x_{\b )\bd} = {\bar D}_{(\ad} \x_{\b \bd )}=0~,
\label{msc}
\ee
which leads to the standard conformal Killing equation
\be
\pa_a \x_b + \pa_b \x_a = \hf\, \eta_{ab}\, \pa_c \x^c\;.
\ee 
The general solution to eq.~(\ref{msc}) was given 
in \cite{Ideas} for $\cN=1$ and in \cite{Park} for $\cN >1$. These conformal Killing supervector fields span a Lie superalgebra which is isomorphic to $\rm su(2,2|1)$.
For the purpose of this paper, it suffices to consider the relation
\be \label{zD}
[\x \;,\; D_\a ] = - (D_\a \x^\b) D_\b
= \hat{\o}_\a{}^\b (z ) D_\b - 
\left(  2 {\bar \s}(z) - \s (z)  \right) D_\a~.
\ee
The superfield parameters $\hat{\o}_{\a \b}(z)$ and $\s (z)$
are expressed in terms of $\x^A = (\x^a, \x^{\a}, \bar \x_{\ad})$ as follows
\be \label{z-dep}
\begin{aligned}
&\hat{\o}_{\a \b}(z) = -D_{(\a} \x_{\b)}\;,
\qquad \s (z) = \frac{1}{6}
\big(  D_\a \x^\a + 2
{\bar D}^\ad {\bar \x}_{\ad} \big)
\end{aligned}
\ee
and can be explicitly found using the components of the conformal Killing supervector, see Refs.~\cite{Ideas, Park, KT} for detail. 
Due to their action on the covariant derivative \eqref{zD}, the $z$-dependent parameters $\hat{\o}_{\a \b}(z)$ and $\s (z)$ can be thought of as the parameters 
of special local Lorentz and scale transformations, respectively. 
These $z$-dependent parameters, along with $\x$, appear in the superconformal transformation law of a primary tensor superfield, see subsection \ref{subsect2.4}.


\subsection{Two-point structures}


Let $z_1$ and $z_2$ be two different points in superspace. In 4D superconformal theories, all building blocks for the two- and three-point correlation 
functions are composed of the two-point structures:
\bsubeq
\bea
x^a_{\bar{1} 2} &=& -x^{a}_{2 \bar{1}} = x^{a}_{1-} - x^a_{2+} + 2 \ri \,\q_{2 }\s^{a} {\bar \q}_{1}~,\\
\q_{12} &=& \q_1 - \q_2~, \qquad {\bar \q}_{12} = {\bar \q}_1 - {\bar \q}_2~,
\eea 
\esubeq
where $x^a_{\pm}= x^a \pm \ri \theta \sigma^a {\bar \theta}$.
In spinor notation, we write
\bsubeq 
\bea
({x}_{\bar{1} 2})^{\ad \a} &=& (\tilde{\s_a})^{\ad \a}x^{a}_{\bar{1} 2}~, \\
({x}_{2 \bar{1}})_{\a \ad} &=& (\s_a)_{\a \ad} \, x^a_{2 \bar{1} }=
-(\s_a)_{\a \ad} \, x^a_{\bar{1} 2}= - \ve_{\a \b} \ve_{\ad \bd} ({x}_{\bar{1} 2})^{\bd \b}~, \label{eps-x}\\
({x}_{\bar{1} 2}^{\,\dagger} )^{\ad \a} &=& - ({x}_{\bar{2} 1})^{\ad \a}~.
\eea
\esubeq
Note that $({x}_{\bar{1} 2})^{\ad \a} (x_{2 \bar{1}})_{\a \bd} = x_{\bar{1} 2}{}^{2}\, \d^{\ad}{}_{\bd}$. 
We sometimes employ matrix-like conventions of \cite{OsbornN1, KT} where the spinor indices are not explicitly written:
\bsubeq
\bea
&&
\psi = (\psi^{\a})~, \quad  \tilde{\psi} = (\psi_{\a})~, \quad  \bar{\psi} = (\bar{\psi}^{\ad})~, \quad \tilde{\bar{\psi}} = (\bar{\psi}_{\ad})~, 
\label{e2a}
\\
&&
x = (x_{\a \ad})~, \quad \tilde{x} = (x^{\ad \a})~.
\label{e2b}
\eea
\esubeq
Since $x^2 \equiv x^a x_a = -\hf {\rm tr}(\tilde{x} x)$, it follows that $\tilde{x}^{-1} = -x/x^2$.
The notation '$\tilde{x}_{\bar{1} 2}$' means that $\tilde{x}_{\bar{1} 2}$ is antichiral with respect to $z_1$ and chiral with respect to $z_2$. 
That is,
\be 
D_{(1)  \a} \tilde{x}_{\bar{1} 2}=0\,, \qquad \bar D_{(2) \bar \a} \tilde{x}_{\bar{1} 2}=0~,
\label{e1}
\ee
where $D_{(1) \a}$ and $\bar D_{(1) \ad}$  are the superspace covariant spinor derivatives acting on the point $z_1$. Similarly,
$D_{(2) \a}$ and $\bar D_{(2) \ad}$ act on the point $z_2$. Explicitly, they are given by 
\bea
D_{\a} = \frac{\pa}{\pa \q^{\a}}+ \ri (\s^a)_{\a \ad} \bar{\q}^{\ad} \frac{\pa}{\pa x^a}~, \qquad \bar D_{\ad} = -\frac{\pa}{\pa \bar \q^{\ad}} - \ri {\q}^{\a} (\s^a)_{\a \ad}  \frac{\pa}{\pa x^a}~.
\eea

The superconformal transformation laws of the two-point structures are given by
\bsubeq
\bea
\d x^{\ad \a}_{{\bar 1}2} &=& -\Big( \hat{{\bar \o}}^\ad{}_\bd (z_1)   
- \d^\ad{}_\bd \,{\bar \s}(z_1)\Big) x^{\bd \a}_{{\bar 1}2}  
- x^{\ad \b}_{{\bar 1}2} \Big( \hat{\o}_\b{}^\a (z_2) -
\d_\b{}^\a \,\s (z_2) \Big) \label{2ptvar-1} \\
\d \q^\a_{12} &=& 2\left({\bar \s}(z_1) - \s (z_1)\right)
\q^\a_{12} - {\rm i} \,\hat{{\bar \eta}}_{\bd}(z_1)
x^{\bd \a}_{{\bar 1}2}
\non \\
&& - \q^\b_{12} 
\left( \hat{\o}_\b{}^\a (z_2) -
\d_\b{}^\a \,\s (z_2) \right)~, 
\label{two-point-variation} 
\eea 
\esubeq
with $\hat{\eta}_\a{} (z) := \hf D_\a \s (z)$.
Let us define
\bea
\cI_{\a \ad}(x_{2 \bar{1}}) = \frac{(x_{2 \bar{1}})_{\a \ad}}{(x_{\bar{1} 2}{}^2)^\hf} \in {\rm SL(2, \mathbb{C})}~. 
\label{I-def}
\eea 
With the aid of \eqref{2ptvar-1}, it follows that $x_{\bar{1} 2}{}^{2}$ and $\cI_{\a \ad}(x_{2 \bar{1}})$ transform covariantly under superconformal 
transformations:
\bsubeq
\bea
\d x_{\bar{1} 2}{}^{2} &=& 2 \big( \bar{\s}(z_1) + \s(z_2) \big)x_{\bar{1} 2}{}^{2}~, \\
\d \cI_{\a \ad}(x_{2 \bar{1}}) &=& \cI_{\a\gd}(x_{2 \bar{1}}) \, \hat{\bar \o}^{\gd}{}_{\ad} (z_1) + \hat{\o}^{\g}{}_{\a} (z_2)\, \cI_{\g \ad}(x_{2 \bar{1}})~. \label{xhat}
\eea
\esubeq

We also note several useful differential identities:
\bsubeq \label{2ptids}
\bea
&&D_{(1) \a} (x_{\bar{2} 1})^{\bd \b} = 4 \ri \d_{\a}{}^{\b} \bar{\q}_{12}^{\bd}~, \qquad \bar{D}_{(1) \ad}\, (x_{\bar{1} 2})^{\bd \b} = 4 \ri \d_{\ad}{}^{\bd} \q_{12}^{\b}~, \\
&&D_{(1) \a} \bigg(\frac{1}{x_{\bar{2} 1}{}^{2}} \bigg) = - \frac{4 \ri}{x_{\bar{2} 1}{}^2}  (\tilde{x}_{\bar{2} 1}{}^{-1})_{\a \bd} \bar{\q}_{12}^{\bd}~, \quad \bar D_{(1) \ad} \bigg(\frac{1}{x_{\bar{1} 2}{}^{2}} \bigg) = - \frac{4 \ri}{x_{\bar{1} 2}{}^2}  (\tilde{x}_{\bar{1} 2}{}^{-1})_{\a \ad} {\q}_{12}^{\a}~,~~~~~
\eea
\esubeq
Here and throughout, we assume that the space points are not coincident, $x_1 \neq x_2$. 


\subsection{Three-point structures}


Given three superspace points $z_1, z_2$ and $z_3$, we have the following three-point structures  $\bm{Z}_1, \bm{Z}_2$ and $\bm{Z}_3$, with $\bm{Z}_1 = (\bm{X}_1^a, \bm{\Q}_1^{\a}, \bm{\bar \Q}_{1}^{\ad})$:
\bsubeq
\bea
&&\bm{X}_{1} = \tilde{x}^{-1}_{\bar{2}1} \tilde{x}_{\bar{2} 3} \tilde{x}^{-1}_{\bar{1} 3}~, \,\,\, \tilde{\bm{\Q}}_{1} = \ri \big(\tilde{x}^{-1}_{\bar{2} 1} \bar{\q}_{12} -  \tilde{x}^{-1}_{\bar{3} 1} \bar{\q}_{13}\big)~, \,\,\, \tilde{\bar{\bm{\Q}}}_{1} = \ri \big({\q}_{12} \tilde{x}^{-1}_{\bar{1} 2} \ - {\q}_{13} \tilde{x}^{-1}_{\bar{1} 3} \big)~,~~~~~ \label{Z1}\\
&&\bm{X}_{2} = \tilde{x}^{-1}_{\bar{3}2} \tilde{x}_{\bar{3} 1} \tilde{x}^{-1}_{\bar{2} 1}~, \,\,\, \tilde{\bm{\Q}}_{2} = \ri \big(\tilde{x}^{-1}_{\bar{3} 2} \bar{\q}_{23} -  \tilde{x}^{-1}_{\bar{1} 2} \bar{\q}_{21}\big)~, \,\,\, \tilde{\bar{\bm{\Q}}}_{2} = \ri \big({\q}_{23} \tilde{x}^{-1}_{\bar{2} 3} \ - {\q}_{21} \tilde{x}^{-1}_{\bar{2} 1} \big)~,~~~~~ \label{Z2}\\
&&\bm{X}_{3} = \tilde{x}^{-1}_{\bar{1}3} \tilde{x}_{\bar{1} 2} \tilde{x}^{-1}_{\bar{3} 2}~, \,\,\, \tilde{\bm{\Q}}_{3} = \ri \big(\tilde{x}^{-1}_{\bar{1} 3} \bar{\q}_{31} -  \tilde{x}^{-1}_{\bar{2} 3} \bar{\q}_{32}\big)~, \,\,\, \tilde{\bar{\bm{\Q}}}_{3} = \ri \big({\q}_{31} \tilde{x}^{-1}_{\bar{3} 1} \ - {\q}_{32} \tilde{x}^{-1}_{\bar{3} 2} \big)~.~~~~~ \label{Z3}
\eea
\esubeq
Since \eqref{Z2} and \eqref{Z3} are obtained through cyclic permutations of superspace points, it suffices to study the properties of \eqref{Z1}. Let us also define 
\bea
\bar{\bm{X}}_{1} = \bm{X}_1^{\dagger} &=& -\tilde{x}_{\bar{3}1}{}^{-1} \tilde{x}_{\bar{3} 2} \tilde{x}_{\bar{1} 2}{}^{-1}~.
\eea
Similar relations hold for $\bar{\bm{X}}_{2}, \bar{\bm{X}}_{3}$.

The structures $\bm{Z}_i$ transform as tensors at the point $z_i,\, i = 1,2,3$. For instance, the transformation law of \eqref{Z1} reads
\bsubeq \label{Z1transf}
\bea
\d \bm{X}_{1\a \ad} &=& \Big( \hat{\o}_{\a}\,^{\b}(z_1) - \d_{\a}\,^{\b}\s(z_1)\Big)  \bm{X}_{1\b \ad} +  \bm{X}_{1\a \bd} \Big( \hat{\bar{\o}}^{\bd}\,_{\ad} (z_1) - \d^{\bd}\,_{\ad} \bar{\s}(z_1)\Big)~,\\
\d {\bm \Q}_{1\, \a} & = & 
\hat{\o}_\a{}^\b (z_1) {\bm \Q}_{1\, \b} 
+ \big( \s (z_1) - 
2{\bar \s} (z_1) \big){\bm \Q}_{1\, \a} ~.
\eea
\esubeq

We list several properties of $\bm{Z}$'s which will be useful later (see \cite{OsbornN1} for details):
\bsubeq \label{XbarX}
\bea
&&\bm{X}_1^2 = \frac{x_{\bar{2} 3}{}^2}{x_{\bar{2} 1}{}^2 x_{\bar{1} 3}{}^2}~, \qquad \bm{\bar X}_1^2 = \frac{x_{\bar{3} 2}{}^2}{x_{\bar{3} 1}{}^2 x_{\bar{1} 2}{}^2}~, \\
&&\bm{\bar X}_{1 \a \ad} = \bm{X}_{1 \a \ad} + \ri \bm{P}_{1 \a \ad}~, \quad \bm{P}_{1 \a \ad} = -4 \bm{\Q}_{1\a} \bm{\bar \Q}_{1\ad}~, \quad \bm{P}_1^2 = -8 \bm{\Q}_1^2 \bm{\bar \Q}_1^2~, \\
&&\frac{1}{\bm{\bar X}_1^2} = \frac{1}{\bm{X}_1^2} -2\ri \frac{(\bm{P}_1 \cdot \bm{X}_1)}{(\bm{X}_1^2)^2}~, \quad (\bm{P}_1 \cdot \bm{X}_1) = -\hf \bm{P}_1^{\ad \a} \bm{X}_{1 \a \ad}~, 
\eea
\esubeq
and hence, $\bm{\bar X}$ is not an independent variable for it can be expressed in terms of 
$\bm{X}, \bm{\Q}, \bm{\bar \Q}$.
The variables $\bm{Z}$ with different labels are related to each other via the identities
\bsubeq \label{Z13}
\bea
\tilde{x}_{\bar{1} 3} \bm{X}_3 \tilde{x}_{\bar{3} 1} &=& - \bar{\bm{X}}_1^{-1} = \frac{\tilde {\bar{ \bm{X}}}_1}{\bar {\bm{X}}_1^2}~, \qquad \tilde{x}_{\bar{1} 3} \bar{\bm{X}}_3 \tilde{x}_{\bar{3} 1} = - \bm{X}_1^{-1} = \frac{\tilde { \bm{X}}_1}{\bm{X}_1^2}~, \\
\frac{x_{\bar{3} 1}{}^2}{x_{\bar{1} 3}{}^2} \tilde{x}_{\bar{1} 3} \tilde{\bm{\Q}}_3 &=&
 - \bm{X}_1^{-1} \tilde{\bm{\Q}}_1~, \qquad \frac{x_{\bar{1} 3}{}^2}{x_{\bar{3} 1}{}^2}  \tilde{\bar {\bm{\Q}}}_3 \tilde{x}_{\bar{3} 1} =
\tilde{\bar{\bm {\Q}}}_1 \bar{\bm{X}}_1^{-1} ~.
\eea
\esubeq
Eqs.~\eqref{Z1transf} and \eqref{Z13} imply that
\bea
\frac{\bm{X}_1^2}{\bm{\bar X}_1^2} = \frac{\bm{X}_2^2}{\bm{\bar X}_2^2} = \frac{\bm{X}_3^2}{\bm{\bar X}_3^2}~,
\eea
and this combination is a superconformal invariant. 


\subsection{Correlation functions of primary superfields}
\label{subsect2.4}


A tensor superfield  $\cO^{\cA}(z)$ is called primary, if it is characterised by the following infinitesimal superconformal transformation law
\be
\begin{aligned}
\d\, \cO^\cA (z) &= - \x \, \cO^\cA (z) 
+ (\hat{\o}^{\a \b} (z) M_{\a \b}+ 
\hat{ \bar{\o}}^{\dot{\a} \dot{\b}} (z) 
\bar{M}_{\dot{\a} \dot{\b}} )^\cA{}_\cB\,
\cO^\cB (z) \\
& \quad - 2\left( q\, \s(z) + \bar{q}\, \bar{\s} (z) \right) 
\cO^\cA(z)~.
\end{aligned}
\label{primary}
\ee
In the above, $\xi$ is the superconformal Killing vector, while $\hat{\o}^{\a \b} (z)$ and $\s(z)$ are the $z$-dependent parameters associated with $\xi$, see eq. \eqref{z-dep}. The superscript `$\cA$' collectively denotes the undotted and dotted spinor indices on which the Lorentz generators $M_{\a \b}$ and $\bar{M}_{\ad \bd}$ act. 
The weights $q$ and $\bar q$ are such that $(q+\bar q)$ is the scale dimension and $(q-\bar q)$ is proportional to the $\rm U(1)$ $R$-symmetry charge of the superfield $\cO^{\cA}$.

Various primary superfields, including conserved current multiplets, are subject to certain differential constraints. These constraints need to be taken into account when computing correlation functions. 
It proves beneficial to make use of these conformally covariant 
operators \cite{OsbornN1}: $\cD_{\bar A} = (\pa / \pa \bX^a_3, \cD_{\a}, \bar{\cD}^{\ad}) $
and $\cQ_{\bar A} = (\pa / \pa \bX^a_3, {\cal Q}_{\a},
\bar{\cal Q}^{\ad})$ defined by 
\be
\begin{aligned}
&\cD_{\a} = \frac{\pa}{ \pa {\bm \Q}^{\a}_3 }
-2{\rm i}\,(\s^a)_{\a \ad} \bar{\bm \Q}^\ad_{3}
\frac{\pa }{ \pa \bX^a_3 }~, \qquad
\bar{\cD}^{\ad} = \frac{\pa}{ \pa \bar{\bm \Q}_{3\, \ad} }~, \\
&{\cal Q}_{\a}  =  \frac{\pa}{ \pa {\bm \Q}^{\a}_3 }~, \qquad
\bar{\cal Q}^{\ad} = \frac{\pa}{ \pa \bar{\bm \Q}_{3 \ad} } 
+ 2{\rm i}\, {\bm \Q}_{3\, \a} (\tilde{\s}^a)^{\ad \a}
\frac{\pa}{ \pa \bX^a_3 }~,  \\
& [\cD_{\bar A}  ,  \cQ_{\bar B} \} ~=~0 ~, 
\end{aligned}
\ee
from which we can derive these anti-commutation relations
\bea
\{ \cD^{\a}, \bar{\cD}^{\ad} \} = 2 \ri (\tilde{\s}^a)^{\ad \a} \frac{\pa}{\pa \bm X^a}~, \qquad \{ {\cQ}^{\a}, \bar{\cQ}^{\ad} \} = -2 \ri\, (\tilde{\s}^a)^{\ad \a} \frac{\pa}{\pa \bm X^a}~.
\eea
One can further prove the following differential identities:
\bsubeq \label{Cderivs}
\bea
D_{(1)\a} \,t(\fn3) &=& -\frac{\ri}{x_{\bar{1} 3}{}^2}(x_{1 \bar{3}})_{\a \ad} \bar{\cD}^{\ad} t(\fn3)~,\\
\bar{D}_{(1)\ad} \,t(\fn3) &=& -\frac{\ri}{x_{\bar{3} 1}{}^2}(x_{3 \bar{1}})_{\a \ad} {\cD}^{\a} t(\fn3)~,\\
D_{(2)\a}\, t(\fn3) &=& \frac{\ri}{x_{\bar{2} 3}{}^2}(x_{2 \bar{3}})_{\a \ad} \bar{\cQ}^{\ad} t(\fn3)~,\\
\bar{D}_{(2) \ad}\, t(\fn3) &=& \frac{\ri}{x_{\bar{3} 2}{}^2}(x_{3 \bar{2}})_{\a \ad} {\cQ}^{\a} t(\fn3)~.
\eea
\esubeq

In accordance with the general prescription of \cite{OsbornN1,Park1, Park},
the two-point function of a primary superfield
$\cO^{\cA}$ with its
conjugate $\bar{\cO}^{\cB}$ is expressed in terms of $\cI$, see eq.~\eqref{I-def}:
\be
\langle \cO^{\cA} (z_1)\;\bar{\cO}^{\cB} (z_2)\rangle
~=~ C_{\cO}\;\frac{ 
\cI^{\cA \cB} ({{x}_{1 \bar{2}}}, {{x}_{2 \bar{1}}}) }
{ (x_{\bar{1}2}{}^2)^{\bar q} (x_{\bar{2}1}{}^2)^q }~,
\label{2pt-gen}
\ee
with $ C_{\cO}$ being a normalisation constant. 

The three-point function 
of primary superfields $\F^{\cA_1}, \J^{\cA_2}$ and $\Pi^{\cA_3}$ has the following general expression \cite{OsbornN1, Park1, Park}:
\bea \label{3ptgen}
&&\langle
\F^{\cA_1} (z_1) \, \J^{\cA_2}(z_2)\,  \Pi^{\cA_3}(z_3)
\rangle 
=\frac{ 
\cI^{\cA_1 \cB_1} ({{x}_{1 \bar{3}}}, {{x}_{3 \bar{1}}})\,
\cI^{\cA_2 \cB_2} ({{x}_{2 \bar{3}}}, {{x}_{3 \bar{2}}})\,
}
{ 
(x_{\bar{1}3}{}^2)^{\bar{q}_1} (x_{\bar{3}1}{}^2)^{q_1} 
(x_{\bar{2}3}{}^2)^{\bar{q}_2} (x_{\bar{3}2}{}^2)^{q_2}
}
H_{\cB_1 \cB_2}{}^{\cA_3} (\fn3)~,~~~~~~~~
\eea 
where the functional form of the tensor $H_{\cB_1 \cB_2}{}^{\cA_3}$ is highly constrained by the superconformal symmetry:
\begin{itemize}

\item[(i)] It possesses the homogeneity property
\be
\begin{aligned}
& H_{\cB_1 \cB_2}{}^{\cA_3} ( \l \bar{\l}\, {\bm X},
\l\, {\bm \Q}, \bar{\l} \bar {\bm \Q}) =
\l^{2a} \bar{\l}^{2\bar{a}}
H_{\cB_1 \cB_2}{}^{\cA_3}( {\bm X}, {\bm \Q}, \bar {\bm \Q})~, \\
& a- 2\bar{a} = \bar{q}_1 + \bar{q}_2 -q_3~,\qquad
\bar{a} - 2a = q_1 + q_2 - \bar{q}_3~,
\end{aligned}
\ee
This condition guarantees that the correlation function has the correct transformation law under the superconformal group. 
By construction, eq.~\eqref{3ptgen} has the correct transformation properties at the points $z_1$ and $z_2$. 
The above homogeneity property implies that it also transforms correctly at $z_3$. The tensor $ H_{\cB_1 \cB_2}{}^{\cA_3}$ has dimension $(a + \bar{a})$.

\item[(ii)]If any of the superfields $\F$, $\J$ and $\P$ obey differential equations (such as conservation laws in the case of conserved current multiplets), then $ H_{\cB_1 \cB_2}{}^{\cA_3}$
is constrained by certain differential equations too. The latter may be derived using \eqref{Cderivs}.

\item[(iii)] If any (or all) of
the superfields $\F$, $\J$ and $\P$ coincide,
then $ H_{\cB_1 \cB_2}{}^{\cA_3}$
obeys certain constraints, as a consequence of the symmetry
under permutations of superspace points. As an example,
\be
\langle \Phi^{{\cal A}}(z_1) \Phi^{{\cal B}}(z_2)
\P^{{\cal C}}(z_3) \rangle =
(-1)^{\epsilon(\Phi)}
\langle \Phi^{{\cal B}}(z_2) \Phi^{{\cal A}}(z_1)
\P^{{\cal C}}(z_3) \rangle~,
\ee
where $\epsilon(\Phi)$ denotes the Grassmann parity of $\Phi^{{\cal A}}$. Note that under permutations of any two superspace points, the three-point building blocks transform as
\bsubeq
\bea
			{\bm X}_{3 \, \a \ad} &\stackrel{1 \leftrightarrow 2}{\longrightarrow} - \bar{\bm X}_{3 \, \a \ad} \, , \hspace{10mm} {\bm \Q}_{3 \, \a} \stackrel{1 \leftrightarrow 2}{\longrightarrow} - {\bm \Q}_{3 \, \a} \, , \label{pt12} \\[2mm]
			{\bm X}_{3 \, \a \ad} &\stackrel{2 \leftrightarrow 3}{\longrightarrow} - \bar{\bm X}_{2 \, \a \ad} \, , \hspace{10mm} {\bm \Q}_{3 \, \a} \stackrel{2 \leftrightarrow 3}{\longrightarrow} - {\bm \Q}_{2 \, \a} \, , \label{pt23} \\[2mm]
			{\bm X}_{3 \, \a \ad} &\stackrel{1 \leftrightarrow 3}{\longrightarrow} - \bar{\bm X}_{1 \, \a \ad} \, , \hspace{10mm} {\bm \Q}_{3 \, \a} \stackrel{1 \leftrightarrow 3}{\longrightarrow} - {\bm \Q}_{1 \, \a} \, . \label{pt13}
\eea
\esubeq
\end{itemize}
The above conditions fix the functional form of $ H_{\cB_1 \cB_2}{}^{\cA_3}$ (and, therefore, the three-point function under consideration) up to a few arbitrary constants.

The goal of this paper is to study how $\cN=1$ superconformal symmetry imposes constraints on the general structure of two- and three-point correlation functions involving a higher-spin conserved spinor current multiplet $S_{\a(k)}$ and its conjugate $\bar S_{\ad(k)}$.  Here the complex superfield $S_{\a(k)} = S_{\a_1 \dots \a_k} = S_{(\a_1 \dots \a_k)}$ is a symmetric rank-$k$ spinor, subject to the following conservation conditions~\cite{KR}
\bsubeq \label{c1}
\bea
D^{\b}S_{\b\a_1 \dots \a_{k-1}} &=& 0 \\
\bar D^2 S_{\a(k)} &=& 0~.
\eea
\esubeq
Making use of the transformation law \eqref{primary}, it can be shown that eqs.~\eqref{c1} are consistent with superconformal invariance provided $S_{\a(k)}$ is a primary superfield with weight $(q, \bar q)= (1+\frac{k}{2}, 1)$ and dimension $2+\frac{k}{2}$. 

Let us first consider the two-point correlation function of $S_{\a(k)}$ and its conjugate $\bar{S}_{\ad(k)}$. Adapting the general prescription \eqref{2pt-gen} to this case leads to
\bea
\la S_{\a(k)} (z_1) \bar{S}_{\ad(k)}(z_2) \ra = \ri^k A \frac{(x_{1 \bar{2}})_{(\a_1(\ad_1} \dots (x_{1 \bar{2}})_{\a_k) \ad_k)}}{x_{\bar{1} 2}{}^2 (x_{\bar{2}1}{}^2)^{k+1} }~,
\label{2ptSS}
\eea
where $A$ is a real coefficient. Using the identities \eqref{2ptids}, along with 
\bea
D_{(1)}^{\a} \bigg[ \frac{(x_{1 \bar{2}})_{\a \ad}}{(x_{\bar{2} 1}{}^2)^{p}}\bigg] = -4 \ri (p-2) \frac{\bar{\q}_{12 \ad}}{(x_{\bar{2} 1}{}^2)^{p}}~,
\eea
one may verify that the correlator \eqref{2ptSS} respects the conservation conditions
\bea
D_{(1)}^{\a_1} \la S_{\a(k)} (z_1) \bar{S}_{\ad(k)}(z_2) \ra = \bar{D}^2_{(1)}\la S_{\a(k)} (z_1) \bar{S}_{\ad(k)}(z_2) \ra = 0~, \non\\
\bar{D}_{(2)}^{\ad_1} \la S_{\a(k)} (z_1) \bar{S}_{\ad(k)}(z_2) \ra = {D}^2_{(2)}\la S_{\a(k)} (z_1) \bar{S}_{\ad(k)}(z_2) \ra = 0~,
\eea
at non-coincident points $z_1 \neq z_2$.
 
In the case of $k=1$, the expression \eqref{2ptSS} is in agreement with \cite{OsbornN1, KT}. Various three-point functions involving $S_{\a}$ and its conjugate $\bar S_{\ad}$ have been studied in detail in \cite{BHK21}.

\section{Correlator $\la \bar S_{\ad(k)}(z_1) S_{\a(k)}(z_2) L(z_3) \ra$}
\label{Section3}


In $\cN=1$ superconformal field theory, the $\rm U(1)$ flavour current multiplet is a primary real superfield $L = \bar{L}$, subject to the conservation equation 
\bea
D^2 L = \bar{D}^2 L = 0~.
\label{flavour-ce}
\eea
Its superconformal transformation law is
\be
\d L = -\xi L -2 (\s + \bar{\s}) L~,
\ee
which means that $L$ has weights $(q, \bar q)= (1, 1)$ and dimension 2. 

In our recent work \cite{BHK21}, we found that the three-point correlator with two spinor current insertions and a flavour current multiplet has two linearly independent functional structures with real coefficients $c_1$ and $d_1$:
\bsubeq \label{JJL-fin}
\bea
\la \bar S_{\ad}(z_1) S_{\b}(z_2) L(z_3) \ra 
= \frac{(x_{3 \bar{1}})_{\ad}{}^{\a} (x_{2 \bar{3}})_{\b}{}^{\bd}}{(x_{\bar{1} 3}{}^2)^{2} x_{\bar{3}1}{}^2 x_{\bar{2} 3}{}^2(x_{\bar{3} 2}{}^2)^{2}} H_{\a \bd}(\fn3)~,
\eea
where
\be
H_{\a \bd} = \ri c_1 \frac{\bm X_{\a \bd}}{\bm X^4}
+ \frac{d_1}{2 \bm X^6} \Big( \bm X^2 \bm P_{\a \bd} - 4 (\bm P \cdot \bm X) \bm X_{\a \bd} \Big)~.
\ee
\esubeq
Here we adopt the notation in which $\bm{X}^k \equiv (\bm{X}^2)^{k/2}$~.

A natural extension is to consider the three-point function $\la \bar S_{\ad(k)}(z_1) S_{\a(k)}(z_2) L(z_3)\ra$, with $k= 1, 2, \dots$.\footnote{We can also consider 
$\la \bar S_{\ad(k)}(z_1) S_{\a(\ell)}(z_2) L(z_3)\ra$ with $k \neq \ell$. However, this correlator carries a non-trivial $R$-symmetry charge 
and, hence, vanishes.}
In accordance with \eqref{3ptgen}, we start with the ansatz
\be
\begin{aligned}
\la \bar S_{\ad(k)}(z_1) S_{\a(k)}(z_2) L(z_3) \ra 
&= \frac{(x_{3 \bar{1}})_{(\ad_1}{}^{\b_1} \dots (x_{3 \bar{1}})_{\ad_k)}{}^{\b_k} (x_{2 \bar{3}})_{(\a_1}{}^{\bd_1} \dots (x_{2 \bar{3}})_{\a_k)}{}^{\bd_k} }{(x_{\bar{1} 3}{}^2)^{k+1} x_{\bar{3}1}{}^2 x_{\bar{2} 3}{}^2(x_{\bar{3} 2}{}^2)^{k+1}}\\
&\times H_{\b_1 \dots \b_k \bd_1 \dots \bd_k}\big(\fn3 \big)~,
\end{aligned}
\label{JJL}
\ee
where the tensor $ H_{\b_1 \dots \b_k \bd_1 \dots \bd_k}$ has the symmetry property $ H_{(\b_1 \dots \b_k) (\bd_1 \dots \bd_k)} = H_{\b(k) \bd(k)}$. It is also characterised by the homogeneity property
\bea
H_{\b(k) \bd (k)}(\l \bar{\l} \bm{X}, \l \bm{\Q}, \bar{\l} \bar{\bm \Q} ) = \l^{-(k+2)} \bar \l^{-(k+2)} H_{\b(k) \bd(k)} (\fxq)~,
\label{scale}
\eea
and hence, its dimension is $-(k+2)$.
Taking the complex conjugate of the correlator \eqref{JJL},
\bea
\la \bar{S}_{\ad(k)} (z_1) S_{\a(k)} (z_2) L(z_3) \ra^{*} = \la \bar{S}_{\ad(k)}(z_2) S_{\a(k)}(z_1) L(z_3) \ra~,
\eea
leads to the reality constraint on $H_{\b(k) \bd(k)}$: 
\be
\bar{H}_{\b(k) \bd(k)}(\fxq) = H_{\b(k) \bd(k)}(-\bm {\bar X}, -{\bm \Q}, -\bar {\bm \Q})~.
\label{re}
\ee
The conservation equations of the higher-spin spinor current and flavour current multiplets result in
\bsubeq
\bea
\bar D^{\ad_1}_{(1)}  \corr1 &=& 0~, \qquad D^2_{(1)}  \corr1 = 0~, \label{ceq1st}\\
D^{\a_1}_{(2)}  \corr1 &=& 0~, \qquad \bar D^2_{(2)}  \corr1 = 0~, \label{ceq2nd}\\
D^2_{(3)} \corr1 &=& 0~,\qquad
\bar D^2_{(3)}  \corr1 = 0~. \label{ceq3rd}
\eea
\esubeq
With the use of identities \eqref{Cderivs}, conditions \eqref{ceq1st} and \eqref{ceq2nd} are translated to 
\bsubeq \label{ceq}
\bea
\cD^{\b_1}H_{\b_1 \dots \b_{k}\bd(k)} = 0~, \label{ceq1}\\
\bar \cD^2 H_{\b(k) \bd(k)} = 0~, \label{ceq2}\\
\bar \cQ^{\bd_1}H_{\b(k) \bd_1 \dots \bd_k} = 0~, \label{ceq3}\\
\cQ^2 H_{\b(k) \bd(k)} = 0~. \label{ceq4}
\eea
\esubeq
Imposing differential constraints \eqref{ceq3rd} is more complicated. We will take care of \eqref{ceq3rd} at the end.  

The problem of computing higher-spin correlator \eqref{JJL} is thus reduced to determining the most general form of $H_{\b(k) \bd(k)}$ satisfying the constraints \eqref{scale}, \eqref{re}, \eqref{ceq3rd} and \eqref{ceq}.

Since $H_{\b(k) \bd(k)}$ is Grassmann even, the most general expansion we can write is
\bea
H_{\b(k) \bd(k)}(\fxq) &=& 
A^{(1)}_{\b(k) \bd(k)} (\bm{X}) + A^{(2)}_{\b(k) \bd(k)} (\bm{X}) {\bm \Q}^2 + A^{(3)}_{\b(k) \bd(k)} (\bm{X}) \bar{\bm \Q}^2 \non\\
&&+ A^{(4)}_{\b(k) \bd(k)} (\bm{X})  {\bm \Q}^2 \bar{\bm \Q}^2 + B_{\b(k) \bd(k), \,\g \dot{\g}} (\bm{X}){\bm \Q}^{\g}  \bar{\bm \Q}^{\gd}\,. 
\label{4.1}
\eea
Constraints \eqref{ceq2} and \eqref{ceq4} immediately tell us that
\be 
A^{(2)}_{\b(k) \bd(k)} = A^{(3)}_{\b(k) \bd(k)} = A^{(4)}_{\b(k) \bd(k)} = 0~,
\label{4.2}
\ee
which leaves us with 
\bsubeq \label{4.3}
\be 
H_{\b(k) \bd(k)}(\fxq) = A_{\b(k) \bd(k)} (\bm{X}) + B_{\b(k) \bd(k), \g \gd} (\bm{X}){\bm \Q}^{\g}  \bar{\bm \Q}^{\dot{\g}}\,,
\ee
where we have redenoted $A_{\b(k) \bd(k)} = A^{(1)}_{\b(k) \bd(k)}$.
Since $A_{\b(k) \bd(k)}$ and $B_{\b(k) \bd(k), \g \gd}$ can be constructed using only tensors $\bm{X}_{ \a \ad}, \ \ve_{\a \b}, \ \ve_{\ad \bd}$, 
it is not hard to list all possible independent structures consistent with the homogeneity property \eqref{scale}
by performing irreducible decompositions into symmetric and antisymmetric parts in dotted and undotted indiices:
\bea
A_{\b(k) \bd(k)} &=& \frac{a_1}{\bm{X}^{2k+2}}\bm{X}_{(\b_1(\bd_1} \dots \bm{X}_{\b_k) \bd_k)}~, \\
B_{\b(k) \bd(k), \g \gd} &=& \frac{b_1}{\bm{X}^{2k+4}} \bm{X}_{(\g(\gd} \bm{X}_{\b_1 \bd_1} \dots \bm{X}_{\b_k) \bd_k)} \non\\
&&+ \frac{b_2}{\bm{X}^{2k+2}} \ve_{\g (\b_1} \ve_{\gd (\bd_1} \bm{X}_{\b_2 \bd_2} \dots \bm{X}_{\b_k) \bd_k)}~.
\label{4.4}
\eea
\esubeq
Here $a_1, b_1, b_2$ are arbitrary complex coefficients. To continue
it is convenient to introduce auxiliary commuting complex variables $(u^{\a}, \bar{w}^{\ad})$, with the property $u^{\a}u_{\a} = \bar{w}_{\ad} \bar{w}^{\ad} = 0$.  
Given a tensor superfield $T_{\a(k) \ad(l)}$, we can associate to it the following index-free superfield 
\bea
T_{(k,l)}(u, \bar{w}) := u^{\a_1} \dots u^{\a_k} \bar{w}^{\ad_1} \dots \bar{w}^{\ad_l} T_{\a_1 \dots \a_k \ad_1 \dots \ad_l}~,
\eea
which is homogeneous of degree $(k,l)$ in the variables $u^{\a}, \bar{w}^{\ad}$. 
Making use of these auxiliary variables and their corresponding partial derivatives $(\pa/\pa u^{\a}, \pa/\pa \bar{w}^{\ad})$ allows us to also convert the conformally covariant derivatives into index-free operators, for example:
\bea
\cD_{(-1,0)} &:=& \cD^{\a}\frac{\pa}{\pa u^{\a}}~, \qquad \cD_{(-1,0)}^2 = 0~, \\
\bar \cQ_{(0,-1)} &:=& \bar \cQ^{\ad} \frac{\pa}{\pa \bar{w}^{\ad}}~, \qquad \bar \cQ_{(0,-1)}^2 = 0~.
\eea 
These nilpotent operators decrease the degree of
homogeneity in $u^{\a}$ and $\bar{w}^{\ad}$.

Using the notation introduced, eq. \eqref{4.3} turns into
\bea
&&H_{(k,k)} (\fxq; \, u, \bar{w}) 
= \frac{a_1}{\bm{X}^{2k+2}} \bm{X}^k_{(1,1)} 
+ \frac{1}{2(k+1)} \bigg\{ \big( b_1 - (k+1) b_2 \big) \frac{(\bm{P} \cdot \bm{X})\bm{X}^k_{(1,1)}}{\bm{X}^{2k+4}} \non\\
&&\qquad \qquad \qquad \qquad \qquad + \big( kb_1 + (k+1)b_2 \big) \frac{(\bm{P} \cdot \bm{\cK})\bm{X}^{k-1}_{(1,1)}}{\bm{X}^{2k+4}}  \bigg\}~,
\label{Htensor-c1}
\eea
where we have defined
\bea
\bm{X}_{(1,1)} &:=& u^{\a} \bar{w}^{\ad} \bm{X}_{\a \ad}~, \qquad \quad \quad \bm{\cK}_{\g \gd} := u^{\a} \bar{w}^{\ad} \bm{X}_{\a \gd} \bm{X}_{\g \ad}~, \label{cK-def}\\
(\bm{P} \cdot \bm{X}) &:=& 2 \bm{\Q}^{\a} \bm{\bar{\Q}}^{\ad} \bm{X}_{\a \ad}~, \qquad (\bm{P} \cdot \bm{\cK}) := 2 \bm{\Q}^{\a} \bm{\bar{\Q}}^{\ad} \bm{\cK}_{\a \ad}~.
\eea
Throughout the paper, we will also often employ the following shorthand notation
\bea
(u \cdot \bm{\Q})&& = u^{\a} \bm{\Q}_{\a}~, \qquad  \qquad \, (\bar{w} \cdot \bar{\bm \Q}) = \bar{w}^{\ad} \bar{\bm \Q}_{\ad}~, \non\\
(\bm{\Q} \bm{X} \bar{w}) &&= \bm{\Q}^{\a} \bm{X}_{\a \ad} \bar{w}^{\ad}~, \qquad  (u \bm{X} \bar{\bm{\Q}} ) = u^{\a} \bm{X}_{\a \ad} \bar{\bm \Q}^{\ad}~.~~~~~
\eea

The differential constraints \eqref{ceq1} and \eqref{ceq3} now read
\bea
\cD_{(-1,0)}H_{(k,k)} &=& 0~, \label{DH-c1} \\
\bar \cQ_{(0,-1)}H_{(k,k)} &=& 0~. \label{QH-c1}
\eea
A useful observation is that 
\bea
\cD_{(-1,0)} \bigg( \frac{\bm{X}^k_{(1,1)}}{\bm{X}^{2k+2}}\bigg) = \bar \cQ_{(0,-1)} \bigg( \frac{\bm{X}^k_{(1,1)}}{\bm{X}^{2k+2}}\bigg) = 0~.
\eea
This means that $a_1$ is an independent coefficient, since the first term in \eqref{Htensor-c1} already satisfies conservation equations.
It can be explicitly checked that 
\bea
\cD_{(-1,0)}H_{(k,k)} =  -(k+1)b_2 \, \bigg\{ (\bar{w} \cdot \bar{\bm{\Q}}) \frac{\bm{X}^{k-1}_{(1,1)}}{\bm{X}^{2k+2}} + 2 \ri \bar{\bm \Q}^2 (\bm{\Q} \bm{X} \bar{w}) \frac{\bm{X}^{k-1}_{(1,1)}}{\bm{X}^{2k+4}} \bigg\}~,
\label{DHkk}
\eea
and
\bea
\bar{\cQ}_{(0,-1)}H_{(k,k)} =  -(k+1)b_2 \, \bigg\{ (u \cdot \bm{\Q}) \frac{\bm{X}^{k-1}_{(1,1)}}{\bm{X}^{2k+2}} + 2 \ri {\bm \Q}^2 (u \bm{X} \bar{\bm{\Q}} ) \frac{\bm{X}^{k-1}_{(1,1)}}{\bm{X}^{2k+4}} \bigg\}~,
\label{QHkk}
\eea
which imply that the conservation laws \eqref{DH-c1} and \eqref{QH-c1} hold provided we set
\bea
b_2 = 0~.
\eea
%
%
In deriving \eqref{DHkk} and \eqref{QHkk}, the following easily verified identities have been used:
\bsubeq \label{ids-DQ}
\bea
&&\cD^{\a} \bm{X}_{\g \gd} = -4\ri \,\d^{\a}{}_{\g} \bar{\bm \Q}_{\gd}~, \qquad \quad \quad ~\bar{\cQ}^{\ad} \bm{X}_{\g \gd} = -4\ri \,\d^{\ad}{}_{\gd} \bm{\Q}_{\g}~,\\
&&\cD^{\a} \bm{X}^k_{(1,1)} = -4 \ri k (\bar{w} \cdot \bar{\bm \Q}) u^{\a} \bm{X}^{k-1}_{(1,1)}~,\\
&&\bar{\cQ}^{\ad} \bm{X}^k_{(1,1)} = -4 \ri k (u \cdot \bm{\Q}) \bar{w}^{\ad} \bm{X}^{k-1}_{(1,1)}~, \\
&&\cD_{(-1,0)} \bm{X}^k_{(1,1)} =  -4 \ri k (k+1) (\bar{w} \cdot \bar{\bm \Q}) \bm{X}^{k-1}_{(1,1)}~, \\
&&\bar \cQ_{(0,-1)} \bm{X}^k_{(1,1)} =  -4 \ri k (k+1) (u \cdot \bm{\Q}) \bm{X}^{k-1}_{(1,1)}~,\\
&&\cD^{\a} \Big( \frac{1}{\bm{X}^{2p}}\Big)= -4 \ri p \frac{\bar{\bm \Q}_{\ad} \bm{X}^{\ad \a}}{\bm{X}^{2p+2}}~, \qquad  \bar \cQ^{\ad} \Big( \frac{1}{\bm{X}^{2p}}\Big)= -4 \ri p \frac{{\bm \Q}_{\a} \bm{X}^{\ad \a}}{\bm{X}^{2p+2}}~,\\
&&\cD^{\a} \bm{P}_{\g \gd} = 4 \,\d^{\a}{}_{\g} \bar{\bm \Q}_{\gd}~, \qquad \qquad \quad \bar{\cQ}^{\ad} \bm{P}_{\g \gd}= 4 \,\d^{\ad}{}_{\gd} \bm{\Q}_{\g}~.
\eea
\esubeq

At this stage, we are left with two unconstrained complex parameters, $a_1$ and $b_1$. 
The next task is to check if the flavour current conservation equations~\eqref{ceq3rd} are satisfied. 
Checking conservation laws on $z_3$ requires more work as there are no identities that would allow differential operators acting on the $z_3$ dependence to pass through the prefactor of \eqref{JJL}. Following the procedure carried out in \cite{BHK21}, let us first express \eqref{JJL} as
\bea
&&
\la \bar S_{\ad(k)}(z_1) S_{\a(k)}(z_2) L(z_3) \ra = \frac{1}{k_1} \cI_{\b(k)\ad(k)}(x_{3 \bar{1}}) \, \cI_{\a(k) \bd(k)}(x_{2 \bar{3}}) H^{\bd(k)\b(k) } (\fn3)~,~~~~~
\label{4.6}
\\
&&
k_1 := (x_{\bar{1} 3}{}^2)^{(k+2)/2} x_{\bar{3} 1}{}^2 x_{\bar{2} 3}{}^2 (x_{\bar{3} 2}{}^2)^{(k+2)/2}~.
\non
\eea
Here the $\cI$-operators are the higher-spin extensions of \eqref{I-def}. Specifically, we define
\bea
\cI_{\a(k) \ad(k)} (x_{3 \bar{1}}) := \frac{1}{(x_{\bar{1} 3}{}^2){}^{k/2}} (x_{3 \bar{1}})_{(\a_1 (\ad_1} \dots (x_{3 \bar{1}})_{\a_k) \ad_k)}~.
\eea
Its inverse will be denoted by
\bea
\bar \cI^{\ad{(k)} \a(k)} (\tilde{x}_{\bar{1} 3}) := \frac{1}{(x_{\bar{1} 3}{}^2){}^{k/2}} (x_{ \bar{1} 3})^{(\ad_1 (\a_1}  \dots (x_{\bar{1}3})^{\ad_k) \a_k)}~.
\eea
These operators satisfy
\bsubeq
\bea
&&\cI_{\a(k) \ad(k)}(x_{3 \bar{1}}) \bar \cI^{\ad{(k)} \g(k)} (\tilde{x}_{\bar{1} 3}) = \d_{(\a_1}^{(\g_1} \dots \d_{\a_k)}^{\g_k)}~,\\
&&\bar \cI^{\gd{(k)} \g(k)} (\tilde{x}_{\bar{1} 3})\cI_{\g(k) \ad(k)}(x_{3 \bar{1}})  = \d_{(\ad_1}^{(\gd_1} \dots \d_{\ad_k)}^{\gd_k)}~.
\eea
\esubeq
It should be kept in mind that the spinor indices can be raised or lowered in accordance with \eqref{eps-x}. This allows us to write
\bea
&&\cI_{\ad(k)}{}^{\a(k)}(x_{3 \bar{1}})= \ve^{\a_1 \b_1} \dots \ve^{\a_k \b_k}\cI_{\b(k) \ad(k)}(x_{3 \bar{1}})~, \\
&&\bar{\cI}_{\a(k)}{}^{\ad(k)}(\tilde{x}_{\bar{1} 3})= \ve_{\a_1 \b_1} \dots \ve_{\a_k \b_k}\bar{\cI}^{\ad(k) \b(k)}(\tilde{x}_{\bar{1} 3})~,
\eea

By rearranging the operators in the three-point function and taking into account their Grassmann parities, one may express the left-hand side of \eqref{4.6} as
\bea \label{JJL-re}
&&
\la \bar S_{\ad(k)}(z_1) S_{\a(k)}(z_2) L(z_3) \ra = (-1)^k \la S_{\a(k)}(z_2) L(z_3) \bar S_{\ad(k)}(z_1) \ra  \non\\
&=& \frac{(-1)^k}{k_2} \cI_{\a(k) \gd(k)}(x_{2 \bar{1}}) \widetilde{H}^{\gd(k)}{}_{\ad(k)}(\bm X_1, \bm \Q_1, \bar {\bm \Q}_1)~, \\
&&
k_2 := (x_{\bar{1} 2}{}^2)^{(k+2)/2} x_{\bar{2} 1}{}^2 x_{\bar{1} 3}{}^2 x_{\bar{3} 1}{}^2~,\non
\eea
for some function $\widetilde{H}^{\gd(k)}{}_{\ad(k)}(\bm X_1, \bm \Q_1, \bar {\bm \Q}_1)$. Comparing eqs.~\eqref{4.6} and~\eqref{JJL-re} allows us to relate $H$ and $\widetilde{H}$
\bea
\widetilde{H}^{\gd(k)}{}_{\ad(k)}(\bm X_1, \bm \Q_1, \bar {\bm \Q}_1) &=& \frac{(-1)^k k_2}{k_1} \bar{\cI}^{\gd(k) \b(k)}
(\tilde{\bar{\bm X}}_1) \cI_{\b(k) \bd(k)} (x_{1 \bar{3}}) \cI_{\a(k) \ad(k)}(x_{3 \bar{1}})\non\\
&&\quad \times H^{\bd(k) \a(k)} (\fn3)~,
\label{tildeHH0}
\eea
where we have used the identity
\bsubeq
\bea
\bar{\cI}^{\ad(k) \a(k)} (\tilde{x}_{\bar{1} 2}) \cI_{\a(k) \bd(k)} (x_{2 \bar{3}}) = 
\bar{\cI}^{\ad(k) \a(k)} (\tilde{\bar{\bm X}}_1) \cI_{\a(k) \bd(k)} (x_{1 \bar{3}})~,
\label{Xinv}
\eea
with 
\bea
\bar{\cI}^{\ad(k) \a(k)} (\tilde{\bar{\bm X}}) = \frac{1}{\bar{\bm X}^k} \bar{\bm{X}}^{ (\ad_1(\a_1} \dots \bar{\bm{X}}^{\ad_k) \a_k)}~.
\eea
\esubeq

In order to compute the expression \eqref{tildeHH0}, we further note that
\bsubeq \label{I-id}
\bea 
&&\cI_{\ad(k)}{}^{\a(k)} (x_{3 \bar{1}}) \bar{\cI}_{\b(k)}{}^{\bd(k)} (\tilde{x}_{\bar{3} 1}) \bm{X}_{3(\a_1(\bd_1} \dots \bm{X}_{3 \a_k) \bd_k)} = (-1)^k \frac{\bar{\bm{X}}_{1 (\b_1(\ad_1} \dots \bar{\bm{X}}_{1 \b_k) \ad_k)}}{\bar{\bm X}_1^{2k} (x_{\bar{1} 3}{}^2 x_{\bar{3} 1}{}^2 )^{k/2}}~,~~~~~~ \\
&&(\bm{P}_3 \cdot \bm{X}_3) = \frac{(\bm{P}_1 \cdot \bm{X}_1)}{\bm{X}_1^2 \bar{\bm{X}}^2_1 x_{\bar{1} 3}{}^2 x_{\bar{3} 1}{}^2}~, \qquad \frac{1}{\bm{X}_{3}^2} = \bar{\bm{X}}^2_1 x_{\bar{1} 3}{}^2 x_{\bar{3} 1}{}^2~.
\eea
\esubeq
Direct computation yields
\bea
&&\widetilde{H}^{\gd(k)}{}_{\ad(k)}(\bm X_1, \bm \Q_1, \bm {\bar \Q}_1) 
= \frac{a_1}{\bm X_1^2} \d_{(\ad_1}^{(\gd_1} \dots \d_{\ad_k)}^{\gd_k)} \non\\
&&\qquad \quad + \frac{b_1}{\bm X_1^4} \Big( 
 \bm{\Q}_1^{\l} \bar{\bm{\Q}}^{\dot{\l}}_{1} \bm{X}_{1 \l \dot{\l}} \d_{(\ad_1}^{(\gd_1} \dots \d_{\ad_k)}^{\gd_k)}
- \frac{k}{k+1} \bm{\Q}_1^{\l} \bar{\bm{\Q}}^{\dot{(\g_1}} \bm{X}_{1 \l (\ad_1} \d^{\gd_2}_{\ad_2} \dots \d^{\gd_k)}_{\ad_k)}~ 
\Big)~,
\eea
which can again be written in index-free notation as
\bea
&&\widetilde{H}(\bm X_1, \bm \Q_1, \bm {\bar \Q}_1; \, \bar{v}, \bar{w}) =  \bar{v}^{\ad_1} \dots \bar{v}^{\ad_k} \bar{w}_{\gd_1} \dots \bar{w}_{\gd_k} \widetilde{H}^{\gd(k)}{}_{\ad(k)}(\bm X_1, \bm \Q_1, \bm {\bar \Q}_1) \non\\
&&= \frac{a_1}{\bm X_1^2} (\bar{v} \cdot \bar{w})^k  + \frac{b_1}{2 \bm X_1^4} (\bar{v} \cdot \bar{w})^{k-1}  \Big\{ (\bm{P}_1 \cdot \bm{X}_1)(\bar{v} \cdot \bar{w}) +\frac{k}{2(k+1)} \bm{P}_1^{\ad \a} \bm{X}_{1 \a \bd} \bar{v}^{\bd} \bar{w}_{\ad}  \Big\}~,~~~~~ \label{tildeHH}
\eea
with $(\bar{v} \cdot \bar{w}) = \ve_{\ad \bd} \bar{v}^{\ad} \bar{w}^{\bd}$.
We now observe that 
\bea
&&
\la S_{\a(k)}(z_2) L(z_3) \bar S_{\ad(k)}(z_1) \ra = [\,\text{relabel} \,z_2\rightarrow z_1, \,z_3 \rightarrow z_2, \,z_1 \rightarrow z_3 ]
\non\\
&&= \la S_{\a(k)}(z_1) L(z_2) \bar S_{\ad(k)}(z_3) \ra
= \frac{1}{k_3} \cI_{\a(k) \gd(k)}(x_{1 \bar{3}}) \widetilde{H}^{\gd(k)}{}_{\ad(k)}(\bm X_3, \bm \Q_3, \bm {\bar \Q}_3)~,
\label{4.7} \\
&&
k_3 := (x_{\bar{3} 1}{}^2)^{(k+2)/2} x_{\bar{3} 2}{}^2 x_{\bar{2} 3}{}^2 x_{\bar{1} 3}{}^2~.
\non
\eea
As a consequence of \eqref{Cderivs}, the conservation conditions on the flavour current multiplet are equivalent to
\bea
\cQ^2 \widetilde{H}(\bm X_3, \bm \Q_3, \bm {\bar \Q}_3; \, \bar{v}, \bar{w})= 0~, 
\qquad \bar \cQ^2 \widetilde{H}(\bm X_3, \bm \Q_3, \bm {\bar \Q}_3; \, \bar{v}, \bar{w})=0~.
\label{4.8}
\eea
It is straightforward to show that \eqref{tildeHH} indeed satisfies~\eqref{4.8} for arbitrary $a_1$ and $b_1$. 

The final step is to impose the reality constraint, \eqref{re}, which requires
\bea
&&\bar{a}_1 = (-1)^k a_1~, \qquad \bar{b}_1 = (-1)^{k+1} b_1~, \non\\
&& \implies a_1 = \ri^k A~, \qquad b_1 = \ri^{k-1} B~,
\eea
where $A$ and $B$ are real. Thus, the higher-spin correlator \eqref{JJL} is fixed by the $\cN=1$ superconformal symmetry up to two independent, real coefficients $A$ and $B$.

As a result, the final form of \eqref{JJL} proves to be
\bsubeq
\bea
&&\la \bar S_{\ad(k)}(z_1) S_{\a(k)}(z_2) L(z_3) \ra 
= \frac{(x_{3 \bar{1}})_{(\ad_1}{}^{\b_1} \dots (x_{3 \bar{1}})_{\ad_k)}{}^{\b_k} (x_{2 \bar{3}})_{(\a_1}{}^{\bd_1} \dots (x_{2 \bar{3}})_{\a_k)}{}^{\bd_k} }{(x_{\bar{1} 3}{}^2)^{k+1} x_{\bar{3}1}{}^2 x_{\bar{2} 3}{}^2(x_{\bar{3} 2}{}^2)^{k+1}}\non\\
&&\hspace{5cm} \times H_{\b_1 \dots \b_k \bd_1 \dots \bd_k}\big(\fn3 \big)~,
\label{342a}
\eea
where the tensor $H_{\b_1 \dots \b_k \bd_1 \dots \bd_k}\big(\fn3 \big)$ has the compact expression
\bea
&&H_{(k,k)} (\fxq; \, u, \bar{w}) 
= u^{\b_1} \dots u^{\b_k} \bar{w}^{\bd_1} \dots \bar{w}^{\bd_k} H_{\b(k) \bd(k)} (\fxq) \non\\
&&= \ri^{k-1} \bigg( \ri A \, \frac{\bm{X}^k_{(1,1)} }{\bm{X}^{2k+2}} 
+ \frac{B}{2(k+1)}  \bigg\{ \frac{(\bm{P} \cdot \bm{X})\bm{X}^k_{(1,1)}}{\bm{X}^{2k+4}} + k \frac{(\bm{P} \cdot \bm{\cK})\bm{X}^{k-1}_{(1,1)}}{\bm{X}^{2k+4}}  \bigg\} \bigg)~.
\eea
\esubeq
Setting $k=1$, direct comparison with \eqref{JJL-fin} shows that we have an agreement provided the parameters are related in the following way:
\be
c_1 = A~, \qquad d_1 = -\frac{B}{4}~.
\ee

In a more general setting, one may also consider
\bea
\la \bar{S}^{\prime}_{\ad(k)} (z_1) S_{\a(k)} (z_2) L(z_3) \ra~,
\label{complex-c1}
\eea
in which case the only difference with \eqref{342a} is that the superfield $\bar{S}^{\prime}_{\ad(k)}$ is not the complex conjugate of $S_{\a(k)}$, i.e. $ (S_{\a(k)})^{*} \neq \bar{S}^{\prime}_{\ad(k)}$. It still obeys the conservation conditions $\bar D^{\ad_1} \bar{S}'_{\ad_1 \dots \ad_k}= D^2 \bar{S}'_{\ad(k)} = 0 $. Since the reality constraint \eqref{re} no longer holds, the correlator \eqref{complex-c1} is determined up to two \textit{complex} coefficients, $a_1$ and $b_1$.


\section{Correlator $\la \bar S_{\ad(k)}(z_1) S_{\a(k)}(z_2) J_{\g \gd}(z_3) \ra$}  \label{Section4}

We turn to constructing the three-point function of the higher-spin spinor current multiplets with the supercurrent, namely $\cor2$.\footnote{The correlator 
$\la \bar S_{\ad(k)}(z_1) S_{\a(\ell)}(z_2) J_{\g \gd}(z_3) \ra$ with $k \neq \ell$ vanishes since it carries a non-trivial $R$-symmetry charge.}

The $\cN=1$ supercurrent \cite{Ferrara:1974pz} is described by a primary real superfield $J_{\a \ad} = \bar J_{\a \ad}$ obeying the conservation law
\bea
D^{\a} J_{\a \ad} = \bar{D}^{\ad} J_{\a \ad} = 0~. \label{cc2}
\eea
Its superconformal transformation is 
\be
\d J_{\a \ad} = -\xi J_{\a \ad} + (\hat{\o}_{\a}\,^{\b} \d_{\ad}^{\,\bd} + \bar{\hat{\o}}_{\ad}\,^{\bd} \d_{\a}^{\,\b}) J_{\b \bd} -3(\s + \bar{\s}) J_{\a \ad}~.
\ee
As a consequence of the reality constraint and conservation law, $J_{\a \ad}$ has weights $(q, \bar q)= (\frac{3}{2}, \frac{3}{2})$ and dimension $3$.

The three-point function of the type $\la \bar S_{\ad}(z_1) S_{\a}(z_2) J_{\g \gd}(z_3) \ra$ was constructed in \cite{BHK21}. It is determined up to three independent, real coefficients:
\bsubeq  \label{corr2fin-k1}
\bea
\la \bar S_{\ad}(z_1) S_{\a}(z_2) J_{\g \gd}(z_3) \ra &=& \frac{(x_{3 \bar 1})_{\b \ad} (x_{2 \bar 3})_{\a \bd}} {(x_{\bar 1 3}{}^2)^{2} (x_{\bar 3 2}{}^2)^{2} x_{\bar 3 1}{}^2 x_{\bar 2 3}{}^2}
H^{\bd \b}\,_{\g \gd} (\fn3)~,
\eea
with
\bea 
&&H^{\bd \b}\,_{\g \gd} (\bm{X}, \bm{\Q}, \bm{\bar{\Q}}) = \frac{1}{2}(d_2+d_3) \frac{1}{\bm{X}^2} \d^{\b}_{\g} \,\d^{\bd}_{\gd} + \hf (d_3 + d_4) \frac{1}{(\bm{X}^2)^2}\bm{X}^{\bd \b} \bm{X}_{\g \gd}  \non\\
&&\qquad \quad + \hf d_4 \frac{\ri}{(\bm{X}^2)^2} \bm{X}^{\bd \b} \bm{P}_{\g \gd} +  \hf d_3 \frac{\ri}{(\bm{X}^2)^2}  \bm{P}^{\bd \b} \bm{X}_{\g \gd} \non\\
&&\qquad \quad - d_2 \frac{\ri}{(\bm{X}^2)^2} (\bm{P} \cdot \bm{X} ) \d^{\b}_{\g} \,\d^{\bd}_{\gd} - d_3 \frac{\ri}{(\bm{X}^2)^3} (\bm{P} \cdot \bm{X})  \bm{X}^{\bd \b} \bm{X}_{\g \gd}~.
\eea
\esubeq

Let us consider $\cor2$. Using the prescription \eqref{3ptgen}, the correlator takes the form
\bea
\cor2 
&=& \frac{(x_{3 \bar{1}})_{(\ad_1}{}^{\b_1} \dots (x_{3 \bar{1}})_{\ad_k)}{}^{\b_k} (x_{2 \bar{3}})_{(\a_1}{}^{\bd_1} \dots (x_{2 \bar{3}})_{\a_k)}{}^{\bd_k} }{(x_{\bar{1} 3}{}^2)^{k+1} x_{\bar{3}1}{}^2 x_{\bar{2} 3}{}^2(x_{\bar{3} 2}{}^2)^{k+1}}\non\\
&&\times H_{ \b(k) \bd(k), \, \g \gd}
(\fn3) ~. \label{cor2}
\eea
Following similar analysis as in section \ref{Section3}, the tensor $H_{ \b(k) \bd(k), \, \g \gd}$ is required to satisfy these properties:
\begin{itemize}
\item \textbf{Homogeneity}: 
\bea
H_{\b(k) \bd (k), \, \g \gd}(\l \bar{\l} \bm{X}, \l \bm{\Q}, \bar{\l} \bar{\bm \Q} ) = \l^{-(k+1)} \bar \l^{-(k+1)} H_{\b(k) \bd(k), \, \g \gd} (\fxq)~,
\label{hom-c2}
\eea
\item \textbf{Reality}:
\bea
&&\la \bar{S}_{\ad(k)} (z_1) S_{\a(k)} (z_2) J_{\g \gd}(z_3) \ra^{*} = \la \bar{S}_{\ad(k)}(z_2) S_{\a(k)}(z_1) J_{\g \gd}(z_3) \ra~, \non\\
&& \implies
\bar{H}_{\b(k) \bd(k), \g \gd}(\fxq) = H_{\b(k) \bd(k), \g \gd}(-\bm {\bar X}, -{\bm \Q}, -\bar {\bm \Q})~.
\label{re-c2}
\eea
\item \textbf{Conservation conditions}:
\bsubeq \label{cons-eq-c2}
\bea
&&\cD^{\b_1}H_{\b_1 \dots \b_{k}\bd(k), \g \gd} = 0~, \qquad 
\bar \cD^2 H_{\b(k) \bd(k), \g \gd} = 0~, \\
&&\bar \cQ^{\bd_1}H_{\b(k) \bd_1 \dots \bd_k, \g \gd} = 0~, \qquad 
\cQ^2 H_{\b(k) \bd(k), \g \gd} = 0~. 
\eea
\esubeq
\end{itemize}
Furthermore, the conservation equations of the supercurrent, eq.~\eqref{cc2}, demand that the following be satisfied:
\be
D^{\g}_{(3)}  \cor2 = 0~,\qquad
\bar D^{\gd}_{(3)}  \cor2 = 0~.
\label{Jgg}
\ee

First of all, the fact that $H_{\b(k)\bd(k), \g \gd}$ is Grassmann even and obeys the differential constraints $\bar \cD^2 H_{\b(k) \bd(k), \g \gd} = 0$ and $\cQ^2 H_{\b(k) \bd(k), \g \gd} = 0$, imply that it has the general form
\bsubeq \label{Htensor-c2}
\be 
H_{\b(k) \bd(k), \g \gd}(\fxq) = A_{\b(k) \bd(k), \g \gd} (\bm{X}) + B_{\b(k) \bd(k), \g \gd, \d \dot{\d}} (\bm{X}){\bm \Q}^{\d}  \bar{\bm \Q}^{\dot{\d}}~. 
\ee
We then write all possible independent structures consistent with \eqref{hom-c2}. This is done by decomposing $A_{\b(k) \bd(k), \g \gd}$ and $B_{\b(k) \bd(k), \g \gd, \d \dot{\d}}$ in terms of their irreducible components 
\bea
&&A_{\b(k) \bd(k), \g \gd} = \frac{a_1}{\bm{X}^{2k+2}} \bm{X}_{(\g (\gd} \bm{X}_{\b_1\bd_1} \dots \bm{X}_{\b_k) \bd_k)}\non\\
&&+ \frac{a_2}{\bm{X}^{2k}} \ve_{\g (\b_1} \ve_{\gd (\bd_1} \bm{X}_{\b_2 \bd_2} \dots \bm{X}_{\b_k) \bd_k)}~, \\
&&B_{\b(k) \bd(k), \g \gd, \d \dot{\d}} = \frac{b_1}{\bm{X}^{2k+4}} \bm{X}_{(\g(\gd} \bm{X}_{\d \dot{\d}} \bm{X}_{\b_1 \bd_1} \dots \bm{X}_{\b_k) \bd_k)} \non\\
&&+ \frac{b_2}{\bm{X}^{2k+2}} \ve_{\g (\d} \ve_{\gd (\dot{\d}} \bm{X}_{\b_1 \bd_1} \dots \bm{X}_{\b_k) \bd_k)} + \frac{b_3}{\bm{X}^{2k+2}} \ve_{\d (\g} \ve_{\dot{\d} (\gd} \bm{X}_{\b_1 \bd_1} \dots \bm{X}_{\b_k) \bd_k)} \non\\
&& + \frac{b_4}{\bm{X}^{2k+2}} \ve_{\g \d} \ve_{\gd \dot{\d}} \bm{X}_{(\b_1 (\bd_1} \dots \bm{X}_{\b_k) \bd_k)} +  \frac{b_5}{\bm{X}^{2k+2}} \ve_{\d (\g} \ve_{\gd (\dot{\d}} \bm{X}_{\b_1 \bd_1} \dots \bm{X}_{\b_k) \bd_k)} \non\\
&&+ \frac{b_6}{\bm{X}^{2k}} \ve_{\g (\b_1} \ve_{|\d| \b_2} \ve_{\gd(\bd_1} \ve_{|\dot{\d}| \bd_2} \bm{X}_{\b_3 \bd_3} \dots \bm{X}_{\b_k) \bd_k)}~,
\label{AB-c2}
\eea
\esubeq
where $a_1, a_2$ and $b_1, \dots b_6$ are arbitrary complex coefficients.
Before imposing conservation conditions \eqref{cons-eq-c2}, it is useful to contract the indices of $H_{\b(k)\bd(k), \g \gd}$ with auxiliary variables. We thus introduce
\bea 
&&H_{(k,k), \g \gd}(\fxq; \, u, \bar{w}) := u^{\b_1} \dots u^{\b_k} \bar{w}^{\bd_1} \dots \bar{w}^{\bd_k} H_{\b(k)\bd(k), \g \gd} \non\\
&&= \frac{\tilde{a}_1}{\bm{X}^{2k+2}} \bm{X}_{\g \gd} \bm{X}^{k}_{(1,1)} 
+ \frac{\tilde{a}_2}{\bm{X}^{2k+2}} \bm{\cK}_{\g \gd} \bm{X}^{k-1}_{(1,1)}\\
&&+ \frac{\bm{\Q}^{\d} \bar{\bm{\Q}}^{\dot{\d}}}{\bm{X}^{2k+4}} 
\Big\{ 
\tilde{b}_1 \bm{X}_{\g \gd} \bm{X}_{\d \dot{\d}} \bm{X}_{(1,1)}^k 
+ \tilde{b}_2 \bm{X}_{\g \dot{\d}} \bm{X}_{\d \dot{\g}} \bm{X}_{(1,1)}^k 
+ \tilde{b}_3 \bm{\cK}_{\g \gd} \bm{X}_{\d \dot{\d}} \bm{X}_{(1,1)}^{k-1} \non\\
&&\qquad \qquad 
+\tilde{b}_4 \bm{\cK}_{\d \dot{\d}} \bm{X}_{\g \dot{\g}} \bm{X}_{(1,1)}^{k-1} 
+ \tilde{b}_5 \bm{\cK}_{\d \gd} \bm{X}_{\g \dot{\d}} \bm{X}_{(1,1)}^{k-1} 
+ \tilde{b}_6 \bm{\cK}_{\g \gd} \bm{\cK}_{\d \dot{\d}} \bm{X}_{(1,1)}^{k-2}\Big\}~,\non
\label{H-c2-compact}
\eea
where $\bm{\cK}_{\g \gd}$ is defined the same way as in \eqref{cK-def}, $\bm{\cK}_{\g \gd} = u^{\a} \bar{w}^{\ad} \bm{X}_{\a \gd} \bm{X}_{\g \ad}$.
In this basis, the (tilde) coefficients are related to those in \eqref{AB-c2} by the rule
\bsubeq \label{ABtilde}
\bea
&&\tilde{a}_1 = \frac{1}{k+1}a_1-a_2~, \qquad \tilde{a}_2 = \frac{k}{k+1}a_1 +a_2~, \\
&&\tilde{b}_1 = -\frac{k-1}{(k+1)(k+2)}b_1 - \frac{2k+1}{(k+1)^2} (b_2+ b_3)-b_4 + \frac{k^2+k+1}{(k+1)^2}b_5 + b_6~, \\
&&\tilde{b}_2 = \frac{1}{k+2}b_1 +\frac{1}{k+1} (b_2+ b_3)+b_4 -b_5~,\\
&&\tilde{b}_3 = \frac{2k}{(k+1)(k+2)}b_1 +\frac{2k}{(k+1)^2} b_2-\frac{k^2}{(k+1)^2}(b_3+b_5)-b_6~,\\
&&\tilde{b}_4 = \frac{2k}{(k+1)(k+2)}b_1 -\frac{k^2}{(k+1)^2}b_2+ \frac{2k}{(k+1)^2} b_3- \frac{k^2}{(k+1)^2}b_5-b_6~,\\
&&\tilde{b}_5 = \frac{k(k+2)}{(k+1)^2}b_5~,\\
&&\tilde{b}_6 = \frac{k(k-1)}{(k+1)^2} \Big( \frac{k+1}{k+2}b_1 + b_2+b_3+b_5 \Big)+b_6~.
\eea
\esubeq

As in the previous section, the first order constraints in \eqref{cons-eq-c2} are equivalent to
\bea
\cD_{(-1,0)} H_{(k,k),\, \g \gd} = 0~, \qquad \bar{\cQ}_{(0,-1)} H_{(k,k),\, \g \gd} = 0~.
\eea
Let us first look at the condition $\cD_{(-1,0)} H_{(k,k),\, \g \gd} = 0$. With the help of identities \eqref{ids-DQ} and considering only the terms linear in $\bar {\bm{\Q}}_{\ad}$, we find the following constraints on the coefficients
\bea
&&k \tilde{a}_1 + \tilde{a}_2 = -\frac{\ri}{4} (k \tilde{b}_1 + \tilde{b}_3 -\tilde{b}_4 + \tilde{b}_6)~, 
\qquad k \tilde{a}_1 = -\frac{\ri}{4} (-k \tilde{b}_2 + \tilde{b}_3 + \tilde{b}_5 + \tilde{b}_6)~,\non\\
&&\qquad \qquad \qquad \qquad (k-1) \tilde{a}_2 = -\frac{\ri}{4} ((k-1) \tilde{b}_3 -2 \tilde{b}_6)~.
\label{E1}
\eea
After some lengthy calculations, it can be shown that the terms cubic in the Grassmann variables (that is, $\sim \bar{\bm \Q}^2 \bm{\Q}^{\d}$) result in
\bea
k (\tilde{b}_1+ \tilde{b}_2)+ \tilde{b}_3 + \tilde{b}_5 + \tilde{b}_6 = 0~, \qquad (k-1) \big( 2 \tilde{b}_3 + \tilde{b}_4 +2 \tilde{b}_5 \big) + (k-3) \tilde{b}_6 = 0~. 
\label{E2}
\eea
In a similar way, one can compute $\bar{\cQ}_{(0,-1)} H_{(k,k),\, \g \gd}$ and show that the terms linear in $\bm{\Q}$ lead to the requirements
\bea
&&k \tilde{a}_1 + \tilde{a}_2 = -\frac{\ri}{4} (k \tilde{b}_1 + \tilde{b}_3 -\tilde{b}_4 -k \tilde{b}_5 + \tilde{b}_6)~, \qquad k \tilde{a}_1 = -\frac{\ri}{4} (-k \tilde{b}_2 + \tilde{b}_3 -k \tilde{b}_5 + \tilde{b}_6)~,\non\\
&&\qquad \qquad \qquad \qquad (k-1) \tilde{a}_2 = -\frac{\ri}{4} \big( (k-1) (\tilde{b}_3 + \tilde{b}_5 )- 2 \tilde{b}_6 \big)~,
\label{E3}
\eea
while the terms proportional to $\bm{\Q}^2 \bar{\bm \Q}^{\dot{\d}}$ give
\bea
k (\tilde{b}_1+ \tilde{b}_2)+ \tilde{b}_3  + \tilde{b}_6 = 0~, \qquad (k-1) \big( 2 \tilde{b}_3 + \tilde{b}_4 + \tilde{b}_5 \big) + (k-3) \tilde{b}_6 = 0~. 
\label{E4}
\eea
The system of equations \eqref{E1}--\eqref{E4} turn out to be consistent and can be solved in terms of three independent coefficients which we choose to be $\tilde{a}_2, \tilde{b}_1$ and $\tilde{b}_3$:
\bea
&&\tilde{b}_5 = 0~, \non\\
&&\tilde{a}_1 = -\frac{1}{4k} \Big( 4(k-1) \tilde{a}_2 + \ri \big( k\tilde{b}_1 + (k+1) \tilde{b}_3 \big) \Big)~, \label{zh1} \\
&&\tilde{b}_2 = \frac{1}{2k} \Big( 4 \ri (k-1) \tilde{a}_2 - 2k \tilde{b}_1 - (k+1) \tilde{b}_3 \Big)~, \qquad \tilde{b}_4 = 2 \ri (k-3) \tilde{a}_2 - \hf (k+1) \tilde{b}_3~, \non\\
&&\tilde{b}_6 = \hf(k-1) (\tilde{b}_3 -4 \ri \tilde{a}_2)~.\non
\eea
Keeping in mind the relations \eqref{ABtilde}, 
one may rewrite~\eqref{zh1} in the basis \eqref{Htensor-c2}. Choosing the three independent coefficients to be $b_1, b_2, b_4$, we find that
\bea \label{coeff-c2}
&&b_5 =0~, \qquad b_6 = 0~, \non\\
&&a_1 = \frac{\ri}{4} \Big( \frac{k+3}{k+1} b_1 + \frac{k+2}{(k+1)^2} b_2 + b_4\Big)~, \qquad a_2 = -\frac{\ri k}{4(k+1)} b_4 ~, \\
&&b_3 = \frac{k+3}{k+2}b_1 + \frac{1}{k+1} b_2~.\non
\eea

It remains to impose conservation equations on the supercurrent $J_{\g \gd}$, eq.~\eqref{Jgg}. As in section \ref{Section3}, this can be done by appropriately rearranging the operators in the correlator and transforming the tensor $H_{ \b(k) \bd(k), \, \g \gd}$. Let us rewrite \eqref{cor2} in terms of the $\cI$-operators:
\bea 
&&\cor2 =
\frac{1}{k_1} \cI_{\b(k) \ad(k)}(x_{3 \bar{1}}) \cI_{\a(k) \bd(k)} (x_{2 \bar{3}}) H^{\bd(k) \b(k)}\,_{\g \gd} (\fn3)~,\non\\
&&k_1: = (x_{\bar{1} 3}{}^2)^{(k+2)/2} (x_{\bar{3} 2}{}^2)^{(k+2)/2} x_{\bar{3} 1}{}^2 x_{\bar{2} 3}{}^2~.
\eea
On the other hand, the same correlator can be written as
\bea
&&\cor2 
= (-1)^k \la {S}_{\a(k)} (z_2) J_{\g \gd} (z_3) \bar{S}_{\ad(k)} (z_1) \ra \non\\
&&= \frac{(-1)^k}{k_2} \cI_{\a(k) \dot{\rho}(k)}(x_{2 \bar{1}}) \cI_{\g \dot{\d}} (x_{3 \bar{1}}) \cI_{\d \dot{\g}}(x_{1 \bar{3}})\widetilde{H}^{\dot{\rho}(k),\, \dot{\d} \d}{}_{\ad(k)}(\bm{X}_1, \bm{\Q}_1, \bar{\bm{\Q}}_1)~,
\label{corr1-tr} \\
&&k_2 := (x_{\bar{1} 2}{}^2)^{(k+2)/2} (x_{\bar{1} 3}{}^2)^{(k+2)/2} (x_{\bar{3} 1}{}^2)^{(k+2)/2} x_{\bar{2} 1}{}^2~.
\non
\eea
Thus, knowing $H$, the tensor $\tilde{H}$ can be computed using the formula
\bea
&&\widetilde{H}^{\dot{\mu}(k)}{}_{\ad(k), \, \d \dot{\d}}\, (\bm{X}_1, \bm{\Q}_1, \bar{\bm{\Q}}_1) 
\\
&&
 = -\frac{k_2}{k_1} \, \bar{\cI}^{\dot{\mu}(k) \mu(k)}
(\tilde{\bar{\bm{X}}}_1)\Big[ \cI_{\ad(k)}{}^{\b(k)}(x_{3 \bar{1}}) \bar{\cI}_{\mu(k)}{}^{\bd(k)} (\tilde{x}_{\bar{3} 1}) {\cI}_{\dot{\d}}{}^{\g} ({x}_{3 \bar{1} }) \bar{\cI}_{\d}{}^{\gd} (\tilde{x}_{\bar{3} 1}) \Big]\non\\
&& \quad \times H_{\b(k)\bd(k), \,\g \gd} (\fn3)~,\non
\label{tH}
\eea
where the identity \eqref{Xinv} has been used.

We now substitute the explicit form of $H_{\b(k)\bd(k), \,\g \gd}$ into \eqref{tH}. This is given by \eqref{Htensor-c2}, but with $b_5 = b_6 = 0$, in accordance with the constraints \eqref{coeff-c2}.
After some tedious calculations and making use of \eqref{I-id}, along with the identities
\bsubeq \label{I-id2}
\bea
&&
\cI_{\a \ad}(x_{1 \bar{3}}) \bar{\bm \Q}_3^{\ad} = \bigg( \frac{x_{\bar{3} 1}{}^2}{\bar{\bm X}_1^2} \bigg)^\hf \frac{\bm{\bar \Q}_{1 \a}^I}{x_{\bar{1} 3}{}^2}~,
\qquad 
\bm{\Q}_3^{\a} \cI_{\a \ad}(x_{3 \bar{1}}) = \bigg( \frac{x_{\bar{1} 3}{}^2}{\bm{X}_1^2} \bigg)^\hf \frac{\bm{\Q}_{1 \ad}^I}{x_{\bar{3} 1}{}^2}~,\\
&&
\bm{\Q}_{1 \ad}^{I} := \bm{\Q}_1^{\a} \cI_{\a \ad}(-\bm{X}_1)  =- \frac{\bm{\Q}_1^{\a} \bm{X}_{1 \a \ad}}{(\bm{X}_1^2)^\hf}~,  \quad
\bar{\bm \Q}_{1 \a}^{I} := \cI_{\a \ad}(\bar{\bm X}_1) \bar{\bm \Q}_1^{\ad} = \frac{\bar{\bm X}_{1 \a \ad}}{(\bar{\bm X}_1^2)^\hf} \bar{\bm \Q}_1^{\ad}~,~~~~~
\eea
\esubeq
we obtain the following expression
\bea 
&&\widetilde{H}_{\d \dot{\d}}(\bm X_1, \bm \Q_1, \bm {\bar \Q}_1; \, \bar{v}, \bar{w}) 
= \bar{v}^{\ad_1} \dots \bar{v}^{\ad_k} \bar{w}_{\dot{\m}_1} \dots \bar{w}_{\dot{\m}_k} \widetilde{H}^{\dot{\m}(k)}{}_{\ad(k),\, \d \dot{\d}}(\bm X_1, \bm \Q_1, \bm {\bar \Q}_1) \non\\
&&= \frac{(\bar{v} \cdot \bar{w})^{k-2}}{4(k+1)^2} \,  \bigg\{ \ri \Big[(k+1)(k+3)b_1 + (k+2) b_2 + (k+1)^2 b_4 \Big] (\bar{v} \cdot \bar{w})^2\, \frac{\bm{X}_{1\d \dot{\d}}}{\bm X_1^4} \non\\
&& + \ri \Big[ k(k+3)b_1+ \frac{k(k+2)}{k+1} b_2 \Big] (\bar{v} \cdot \bar{w}) \, \frac{\bm{\cV}_{1\d \dot{\d}}}{\bm X_1^4} 
+ \Big[(k-1)b_1+ \frac{k(k+2)}{k+1} b_2 \Big] (\bar{v} \cdot \bar{w})^2\, \frac{\bm{P}_{1\d \dot{\d}}}{\bm X_1^4} \non\\
&&+ 4(k+1)b_1 (\bar{v} \cdot \bar{w})^2\, \frac{(\bm{P}_1 \cdot \bm{X}_1)\bm{X}_{1\d \dot{\d}}}{\bm{X}_1^6} 
+ \Big[ 8k b_1 + \frac{4k(k+2)}{k+1} b_2 \Big](\bar{v} \cdot \bar{w})\, \frac{(\bm{P}_1 \cdot \bm{X}_1)\bm{\cV}_{1\d \dot{\d}}}{\bm{X}_1^6}\non\\
&&- \Big[ 2k(k-1) b_1 - \frac{2k(k+2)}{k+1} b_2 \Big](\bar{v} \cdot \bar{w})\, \frac{(\bm{P}_1 \cdot \bm{\cV}_1)\bm{X}_{1\d \dot{\d}}}{\bm{X}_1^6} \non\\
&&+ \Big[ k(k+3) b_1 + \frac{k(k+2)}{k+1}b_2\Big](\bar{v} \cdot \bar{w})\, \frac{\bm{P}_1^{\gd \g} \bm{\cV}_{1 \d \gd} \bm{X}_{1 \g \dot{\d}}}{\bm{X}_1^6}\non\\
&&- \Big[ 4k(k-1) b_1 + \frac{2k(k-1)(k+2)}{k+1}b_2\Big] \frac{(\bm{P}_1 \cdot \bm{\cV}_1) \bm{\cV}_{1 \d \dot{\d}}}{\bm{X}_1^6} \bigg\}~.
\label{tH-c2-vw}
\eea
We recall that $b_1, b_2$ and $b_4$ are independent. We have also defined
$
\bm{\cV}_{\a \ad} := \bm{X}_{\a \bd} \, \bar{v}_{\ad} \bar{w}^{\bd}~.
$

Going back to eq.~\eqref{corr1-tr} and relabelling superspace points, we see that
\bea
&&
\la S_{\a(k)}(z_2) J_{\g \gd}(z_3) \bar S_{\ad(k)}(z_1) \ra = [\,\text{relabel} \,z_2\rightarrow z_1, \,z_3 \rightarrow z_2, \,z_1 \rightarrow z_3 ]
\non\\
&&= \la S_{\a(k)}(z_1) J_{\g \gd}(z_2) \bar S_{\ad(k)}(z_3) \ra \\
&&= -\frac{1}{k_3} \cI_{\a(k) \dot{\mu}(k)}(x_{1 \bar{3}}) \bar{\cI}_{\g}{}^{\dot{\d}} (\tilde{x}_{\bar{3} 2}) {\cI}_{\dot{\g}}{}^{\d}({x}_{ 3 \bar{2}})\widetilde{H}^{\dot{\mu}(k)}{}_{\ad(k),\, \d \dot{\d}}(\bm{X}_1, \bm{\Q}_1, \bar{\bm{\Q}}_1)~,\non\\
&&
k_3 := (x_{\bar{3} 1}{}^2)^{(k+2)/2} (x_{\bar{3} 2}{}^2)^{(k+2)/2} (x_{\bar{2} 3}{}^2)^{(k+2)/2} x_{\bar{1} 3}{}^2~.
\non
\eea
By virtue of eq.~\eqref{Cderivs}, the conservation law of $J_{\g \gd}$ requires that the following equations must hold:
\bea
\cQ^{\d}\widetilde{H}_{\d \dot{\d}}(\bm X_3, \bm \Q_3, \bm {\bar \Q}_3; \, \bar{v}, \bar{w}) = 0~, \qquad \bar \cQ^{\dot{\d}}\widetilde{H}_{\d \dot{\d}}(\bm X_3, \bm \Q_3, \bm {\bar \Q}_3; \, \bar{v}, \bar{w}) = 0~. 
\label{427}
\eea
It is not hard to verify that \eqref{tH-c2-vw} satisfies \eqref{427} for an arbitrary choice of complex coefficients $b_1, b_2, b_4$. 

The last step is to impose the reality condition \eqref{re-c2}:
\bea
\bar{H}_{\b(k) \bd(k), \g \gd}(\fxq) = H_{\b(k) \bd(k), \g \gd}(-\bm {\bar X}, -{\bm \Q}, -\bar {\bm \Q})~.
\eea
Here $\bar{H}_{\b(k) \bd(k), \g \gd}(\fxq)$ means that we are taking the complex conjugate of the expression \eqref{Htensor-c2}. This implies that
\bea
b_1 = \ri^k A~, \qquad b_2 = \ri^k B~, \qquad b_4 = \ri^k C~,
\eea
where $A, B, C$ are real coefficients. 

The correlator \eqref{cor2} is thus fixed up to three real coefficients. It is given by
\bsubeq \label{c2-fin-compact}
\bea
&&\cor2
= \frac{(x_{3 \bar{1}})_{(\ad_1}{}^{\b_1} \dots (x_{3 \bar{1}})_{\ad_k)}{}^{\b_k} (x_{2 \bar{3}})_{(\a_1}{}^{\bd_1} \dots (x_{2 \bar{3}})_{\a_k)}{}^{\bd_k} }{(x_{\bar{1} 3}{}^2)^{k+1} x_{\bar{3}1}{}^2 x_{\bar{2} 3}{}^2(x_{\bar{3} 2}{}^2)^{k+1}}\non\\
&&\hspace{5cm} \times H_{\b(k)\bd(k),\, \g \gd}\big(\fn3 \big)~.
\label{c2-fin-compact-A}
\eea
For convenience, here we rewrite explicitly the form of $H_{\b(k)\bd(k),\, \g \gd}\big(\fn3 \big)$ in the basis \eqref{H-c2-compact}
\bea
&&H_{(k,k), \g \gd}(\fxq; \, u, \bar{w}) := u^{\b_1} \dots u^{\b_k} \bar{w}^{\bd_1} \dots \bar{w}^{\bd_k} H_{\b(k)\bd(k), \g \gd} \non\\
&&= \frac{\tilde{a}_1}{\bm{X}^{2k+2}} \bm{X}_{\g \gd} \bm{X}^{k}_{(1,1)} 
+ \frac{\tilde{a}_2}{\bm{X}^{2k+2}} \bm{\cK}_{\g \gd} \bm{X}^{k-1}_{(1,1)}\\
&&+ \frac{\bm{\Q}^{\d} \bar{\bm{\Q}}^{\dot{\d}}}{\bm{X}^{2k+4}} 
\Big\{ 
\tilde{b}_1 \bm{X}_{\g \gd} \bm{X}_{\d \dot{\d}} \bm{X}_{(1,1)}^k 
+ \tilde{b}_2 \bm{X}_{\g \dot{\d}} \bm{X}_{\d \dot{\g}} \bm{X}_{(1,1)}^k 
+ \tilde{b}_3 \bm{\cK}_{\g \gd} \bm{X}_{\d \dot{\d}} \bm{X}_{(1,1)}^{k-1} \non\\
&&\qquad \qquad 
+\tilde{b}_4 \bm{\cK}_{\d \dot{\d}} \bm{X}_{\g \dot{\g}} \bm{X}_{(1,1)}^{k-1} 
+ \tilde{b}_6 \bm{\cK}_{\g \gd} \bm{\cK}_{\d \dot{\d}} \bm{X}_{(1,1)}^{k-2}\Big\}~\non,
\eea
where
\bea
\tilde{a}_1 &=& \frac{\ri^{k+1}}{4(k+1)^3} \Big( (k+1)(k+3) A + (k+2)B + (k+1)^3 C \Big)~, \non\\
\tilde{a}_2 &=& \frac{\ri^{k+1} k}{4(k+1)^3} \Big( (k+1)(k+3) A + (k+2)B \Big)~, \non\\
\tilde{b}_1 &=& -\frac{\ri^{k}}{(k+1)^3} \Big( (k+1)(3k+1) A + (2k+1)(k+2)B + (k+1)^3 C\Big)~, \non\\
\tilde{b}_2 &=& \frac{\ri^{k}}{(k+1)^2} \Big( 2(k+1)A + (k+2)B + (k+1)^2 C\Big)~,\\
\tilde{b}_3 &=& -\frac{\ri^{k} k}{(k+1)^3} \Big( (k^2-1) A - (k+2)B\Big)~,\non\\
\tilde{b}_4 &=& \frac{\ri^{k}}{(k+1)^3} \Big( 4k(k+1)A-k(k+2)(k-1)B \Big)~,\non\\
\tilde{b}_6 &=& \frac{\ri^{k} k(k-1)}{(k+1)^3} \Big( 2(k+1)A+(k+2)B \Big)~.\non
\eea
\esubeq
Setting $k=1$ and comparing \eqref{c2-fin-compact} with \eqref{corr2fin-k1}, we see that we have an agreement with the parameters related by
\be
d_2 = -\frac{3B}{16}~, \qquad d_3 = -\frac{A}{2}~, \qquad d_4 = -\frac{1}{8} (3B + 4(A+C))~.
\ee

In a more general setting, one may also consider
\bea
\la \bar{S}^{\prime}_{\ad(k)} (z_1) S_{\a(k)} (z_2) J_{\g \gd}(z_3) \ra~.
\label{complex-c2}
\eea
The only difference with \eqref{c2-fin-compact-A} is that now $ (S_{\a(k)})^{*} \neq \bar{S}^{\prime}_{\ad(k)}$. Since one can no longer impose the reality constraint \eqref{re-c2}, the correlator \eqref{complex-c2} is determined up to three \textit{complex} coefficients, i.e. $b_1, b_2$ and $b_4$. 


\section{Correlator $\la \bar S_{\ad(k)}(z_1) S_{\b(k+l)}(z_2) \bar{S}_{\gd(l)} (z_3) \ra$}  \label{Section5}

We consider a three-point function of the higher-spin spinor current multiplets in the form $\c3$.\footnote{The correlator 
$\la \bar S_{\ad(k)}(z_1) S_{\b(m)}(z_2) \bar{S}_{\gd(l)} (z_3) \ra$ with $m \neq k+\ell$ vanishes because it carries a non-trivial 
$R$-symmetry charge.}
Its general form is given by
\bea
\c3
&=& \frac{(x_{3 \bar{1}})_{(\ad_1}{}^{\a_1} \dots (x_{3 \bar{1}})_{\ad_k)}{}^{\a_k} (x_{2 \bar{3}})_{(\b_1}{}^{\bd_1} \dots (x_{2 \bar{3}})_{\b_{k+l})}{}^{\bd_{k+l}} }{(x_{\bar{1} 3}{}^2)^{k+1} x_{\bar{3}1}{}^2 x_{\bar{2} 3}{}^2(x_{\bar{3} 2}{}^2)^{k+l+1}}\non\\
&&\times H_{ \a(k) \bd(k+l), \, \gd(l)}
(\fn3) ~, \label{c3}
\eea
with $k,\, l = 1,2, \dots$. The tensor $H_{ \a(k) \bd(k+l), \, \gd(l)}$ is subject to the following constraints:
\begin{itemize}
\item \textbf{Homogeneity}: 
\bea
H_{\a(k) \bd (k+l), \, \gd(l)}(\l \bar{\l} \bm{X}, \l \bm{\Q}, \bar{\l} \bar{\bm \Q} ) = \l^{-(k+2)} \bar \l^{-(k+2)} H_{\a(k) \bd(k+l), \, \gd(l)} (\fxq)~,
\label{hom-c3}
\eea
\item \textbf{Symmetry under $z_1 \leftrightarrow z_3$}:
\bea
&&\c3 = (-1)^{k+l+kl} \la \bar{S}_{\gd(l)}(z_3) S_{\b(k+l)}(z_2) \bar{S}_{\ad(k)}(z_1) \ra \non\\
&& =  (-1)^{k+l+kl} \la \bar{S}_{\ad(l)}(z_1) S_{\b(k+l)}(z_2) \bar{S}_{\gd(k)}(z_3) \ra \bigg|_{z_1 \leftrightarrow z_3,~\ad \leftrightarrow \gd}~,
\label{inv-c3}
\eea
where we have taken into account the Grassmann parities of the superfields. 
\item \textbf{Conservation conditions}:
\bsubeq \label{cons-eq-c3}
\bea
&&\cD^{\a_1}H_{\a_1 \dots \a_{k}\bd(k+l), \gd(l)} = 0~, \qquad 
\bar \cD^2 H_{\a(k) \bd(k+l), \gd(l)} = 0~, \\
&&\bar \cQ^{\bd_1}H_{\a(k) \bd_1 \dots \bd_{k+l}, \gd(l)} = 0~, \qquad 
\cQ^2 H_{\a(k) \bd(k+l), \gd(l)} = 0~. 
\eea
\esubeq
\end{itemize}
Upon imposing \eqref{cons-eq-c3}, it follows from the symmetry property \eqref{inv-c3} that the conservation conditions at $z_3$ are automatically satisfied,
\be
\bar{D}^{\gd_1}_{(3)}  \c3 = 0~,\quad
D^2_{(3)}  \c3 = 0~.
\ee

The constraints $\bar \cD^2 H_{\a(k) \bd(k+l), \gd(l)} = 0$ and $\cQ^2 H_{\a(k) \bd(k+l), \gd(l)} = 0$ tell us that $H_{\a(k) \bd(k+l), \gd(l)}$, being 
Grassmann even, can be expressed in the form
\bsubeq \label{Htensor-c3}
\be 
H_{\a(k) \bd(k+l), \gd(l)}(\fxq) = A_{\a(k) \bd(k+l), \gd(l)} (\bm{X}) + B_{\a(k) \bd(k+l), \gd(l), \d \dot{\d}} (\bm{X}){\bm \Q}^{\d}  \bar{\bm \Q}^{\dot{\d}}~.
\ee
Let us then contract the indices using auxiliary variables and list all possible independent structures consistent with \eqref{hom-c3}. 
The independent tensor structures can be constructed by performing decomposition of $ A_{\a(k) \bd(k+l), \gd(l)} (\bm{X})$ and 
$B_{\a(k) \bd(k+l), \gd(l), \d \dot{\d}} (\bm{X})$ 
into the irreducible components. We obtain
\bea
&&H_{(k,k+l,l)} (\fxq; \, u, \bar{v}, \bar{w}) \non\\ 
&&= u^{\a_1} \dots u^{\a_k} \bar{w}^{\bd_1} \dots \bar{w}^{\bd_{k+l}} \bar{v}^{\gd_1} \dots \bar{v}^{\gd_l}  \bigg\{
\frac{a_1}{\bm{X}^{2k+2}} \ve_{\gd_1 \bd_1} \dots \ve_{\gd_l \bd_l} \bm{X}_{\a_1 \bd_{l+1}} \dots \bm{X}_{\a_k \bd_{k+l}} \non\\
&&+ \frac{\bm{\Q}^{\d} \bar{\bm \Q}^{\dot{\d}}}{\bm{X}^{2k+2}} 
\Big( 
\frac{b_1}{\bm{X}^2} ~\ve_{\gd_l \dot{\d}} \ve_{\gd_1\bd_1}\dots \ve_{\gd_{l-1} \bd_{l-1}} \bm{X}_{(\d \bd_l} \bm{X}_{\a_1 \bd_{l+1}} \dots \bm{X}_{\a_k) \bd_{k+l}} \non\\
&&\qquad \qquad +\frac{b_2}{\bm{X}^2} ~\ve_{\bd_l \dot{\d}} \ve_{\gd_1\bd_1}\dots \ve_{\gd_{l-1} \bd_{l-1}} \bm{X}_{(\d \gd_l} \bm{X}_{\a_1 \bd_{l+1}} \dots \bm{X}_{\a_k) \bd_{k+l}} \non\\
&&\qquad \qquad +\,  b_3 ~ \ve_{\d \a_1} \ve_{\dot{\d} \bd_1} \ve_{\gd_1 \bd_2}\dots \ve_{\gd_{l} \bd_{l+1}} \bm{X}_{\a_2 \bd_{l+2}} \dots \bm{X}_{\a_k \bd_{k+l}}
\Big) \bigg\} \label{Htensor-c3b} \\
&&= a_1(\bar{v} \cdot \bar{w})^l \frac{\bm{X}^k_{(1,1)}}{\bm{X}^{2k+2}} + b_1  (\bar{v} \cdot \bar{w})^{l-1} (\bm{\Q} \bm{X} \bar{w}) (\bar{v} \cdot \bar{\bm \Q}) \frac{\bm{X}^k_{(1,1)}}{\bm{X}^{2k+4}} \non\\
&&+b_2 (\bar{v} \cdot \bar{w})^{l-1} (\bm{\Q} \bm{X} \bar{v}) (\bar{w} \cdot \bar{\bm \Q}) \frac{\bm{X}^k_{(1,1)}}{\bm{X}^{2k+4}} \non\\
&&+ \Big( b_3 -\frac{k}{k+1} b_2 \Big)(\bar{v} \cdot \bar{w})^{l} (u \cdot \bm{\Q}) (\bar{w} \cdot \bar{\bm \Q}) \frac{\bm{X}^{k-1}_{(1,1)}}{\bm{X}^{2k+2}}~. \non
\eea
\esubeq
All the coefficients above are complex. Looking at the conformally covariant structures in \eqref{Htensor-c3b} after contraction with the auxiliary spinors,
it might appear that there are more admissible structures which can be added to the list, namely
\bea
&&(\bar{v} \cdot \bar{w})^{l} (\bm{\Q} \bm{X} \bar{w}) (u \bm{X} \bar{\bm \Q}) \frac{\bm{X}^{k-1}_{(1,1)}}{\bm{X}^{2k+2}}~, \qquad (\bar{v} \cdot \bar{w})^{l} (\bm{\Q} \bm{X} \bar{\bm \Q}) \frac{\bm{X}^{k}_{(1,1)}}{\bm{X}^{2k+4}}~, \non\\
&& \qquad \qquad 
(\bar{v} \cdot \bar{w})^{l-1} (\bm{\Q} \bm{X} \bar{w}) (\bar{w} \cdot \bar{\bm \Q}) (u \bm{X} \bar{v}) \frac{\bm{X}^{k-1}_{(1,1)}}{\bm{X}^{2k+4}}~.
\eea
However, these structures prove to be linearly dependent of the others. More precisely, one can prove that
\bsubeq
\bea
&&(\bar{v} \cdot \bar{w})^{l} (\bm{\Q} \bm{X} \bar{w}) (u \bm{X} \bar{\bm \Q}) \frac{\bm{X}^{k-1}_{(1,1)}}{\bm{X}^{2k+2}} 
= 
(\bar{v} \cdot \bar{w})^{l-1} \Big[ 
(\bm{\Q} \bm{X} \bar{w}) (\bar{v} \cdot \bar{\bm \Q}) - (\bm{\Q} \bm{X} \bar{v}) (\bar{w} \cdot \bar{\bm \Q}) \Big]
\frac{\bm{X}^k_{(1,1)}}{\bm{X}^{2k+4}} \non\\
&&\qquad \qquad \qquad \qquad \qquad \qquad \qquad + (\bar{v} \cdot \bar{w})^{l} (u \cdot \bm{\Q}) (\bar{w} \cdot \bar{\bm \Q}) \frac{\bm{X}^{k-1}_{(1,1)}}{\bm{X}^{2k+2}}~, \\
&&(\bar{v} \cdot \bar{w})^{l} (\bm{\Q} \bm{X} \bar{\bm \Q}) \frac{\bm{X}^{k}_{(1,1)}}{\bm{X}^{2k+4}}
= 
(\bar{v} \cdot \bar{w})^{l-1}  \Big[ (\bm{\Q} \bm{X} \bar{w}) (\bar{v} \cdot \bar{\bm \Q})  - (\bm{\Q} \bm{X} \bar{v}) (\bar{w} \cdot \bar{\bm \Q}) \Big] \frac{\bm{X}^k_{(1,1)}}{\bm{X}^{2k+4}}~, \\
&&(\bar{v} \cdot \bar{w})^{l-1} (\bm{\Q} \bm{X} \bar{w}) (\bar{w} \cdot \bar{\bm \Q}) (u \bm{X} \bar{v}) \frac{\bm{X}^{k-1}_{(1,1)}}{\bm{X}^{2k+4}}  \non\\
&&= (\bar{v} \cdot \bar{w})^{l-1} \frac{\bm{X}^{k-1}_{(1,1)}}{\bm{X}^{2k+2}} \Big[ (\bm{\Q} \bm{X} \bar{v}) \frac{\bm{X}_{(1,1)} }{\bm{X}^2}-  (\bar{v} \cdot \bar{w}) (u \cdot \bm{\Q}) \Big] (\bar{w} \cdot \bar{\bm \Q}) ~.
\eea
\esubeq

Next, we impose conservation laws \eqref{cons-eq-c3}. The first condition, $\cD^{\a_1} H_{\a_1 \dots \a_k \bd(k+l),\, \gd(l)} = 0$, is equivalent to requiring
\bea
\cD_{(-1,0,0)} H_{(k,k+l, l)} (\fxq;\, u, \bar{v}, \bar{w}) = 0~, \qquad \cD_{(-1,0,0)} := \cD^{\a} \frac{\pa}{\pa u^{\a}}~.
\label{c3-cc1}
\eea
Keeping in mind the relations \eqref{ids-DQ} and further noting that
\bea
\cD_{(-1,0,0)}  \bigg( \frac{\bm{X}^k_{(1,1)}}{\bm{X}^{2k+2}}\bigg) = \cD_{(-1,0,0)} \bigg( (\bm{\Q} \bm{X} \bar{w}) (\bar{v} \cdot \bar{\bm \Q}) \frac{\bm{X}^k_{(1,1)}}{\bm{X}^{2k+4}}\bigg) = 0~,
\eea
it can be easily checked that \eqref{c3-cc1} gives rise to
\bea
b_3 = 0~.
\label{b3-c3}
\eea
The second conservation condition reads 
\bea
\bar {\cQ}_{(0,-1,0)} H_{(k,k+l, l)} = 0~, \qquad \bar \cQ_{(0,-1,0)} := \bar \cQ^{\ad} \frac{\pa}{\pa \bar{w}^{\ad}}~.
\label{c3-cc2}
\eea
Using the fact that $b_3 = 0$, we obtain
\bea
0 &&= \bar \cQ_{(0,-1,0)} H_{(k,k+l,l)} \non\\
&&= \big[ 4 \ri l(k+1) a_1 -(k+1) b_1 - (k+l+1)b_2 \big] \non\\
&&\quad \times  \bigg\{  (\bar{v} \cdot \bar{w} )^{l-1} (\bm{\Q} \bm{X} \bar{v})\frac{\bm{X}^{k}_{(1,1)}}{\bm{X}^{2k+4}} - \frac{k}{k+1} (\bar{v} \cdot \bar{w} )^{l}(u \cdot \bm{\Q}) \frac{\bm{X}^{k-1}_{(1,1)}}{\bm{X}^{2k+2}}  \bigg\} \non\\
&&\implies b_2 = \frac{k+1}{k+l+1} \big( 4\ri \,l a_1 - b_1 \big)~.
\label{b2-c3}
\eea
Imposing conservation laws thus leaves us with only two independent coefficients, which we choose to be $a_1$ and $b_1$. 

We turn to analysing the algebraic constraint \eqref{inv-c3}. Under the exchange $z_1 \leftrightarrow z_3$, the three-point function possesses the property
\bea
&&\c3 = (-1)^{k+l+kl} \la \bar{S}_{\gd(l)}(z_3) S_{\b(k+l)}(z_2) \bar{S}_{\ad(k)}(z_1) \ra~ \non\\
&&= (-1)^{k+l+kl} \la \bar{S}_{\ad(l)}(z_1) S_{\b(k+l)}(z_2) \bar{S}_{\gd(k)}(z_3) \ra \bigg|_{z_1 \leftrightarrow z_3,~\ad \leftrightarrow \gd}~.
\label{inv-c3-2}
\eea
As per usual, upon expressing the first line of \eqref{inv-c3-2} in terms of the $\cI$-operators, we have
\bea \label{5.15}
&&\frac{(-1)^{k+l}}{k_1} \cI_{\ad(k)}{}^{\a(k)} (x_{3 \bar{1}}) \bar{\cI}_{\b(k+l)}{}^{\bd(k+l)} (\tilde{x}_{\bar{3} 2}) H_{\a(k)\bd(k+l), \, \gd(l)} (\fn3) \\
&&= 
(-1)^{kl}\,  \frac{1}{k_2} \cI_{\gd(l)}{}^{\r(l)} (x_{1 \bar{3}}) \bar{\cI}_{\b(k+l)}{}^{\dot{\r}(k+l)} (\tilde{x}_{\bar{1} 2 }) \widetilde{H}_{\r(l)\dot{\r}(k+l), \, \ad(k)} (\bm{X}_1, \bm{\Q}_1, \bar{\bm \Q}_1)~, \non\\
&&k_1: = (x_{\bar{1} 3}{}^2)^{(k+2)/2} x_{\bar{3} 1}{}^2  x_{\bar{2} 3}{}^2 (x_{\bar{3} 2}{}^2)^{(k+l+2)/2} ~, \qquad 
k_2: = x_{\bar{1} 3}{}^2 (x_{\bar{3} 1}{}^2)^{(l+2)/2}  x_{\bar{2} 1}{}^2 (x_{\bar{1} 2}{}^2)^{(k+l+2)/2} ~,~ \non
\eea
which then yields the relation
\bea \label{tH-c3}
&&\widetilde{H}_{\r(l)\dot{\r}(k+l), \,\ad(k)} (\bm{X}_1, \bm{\Q}_1, \bar{\bm \Q}_1) \non\\
&&= (-1)^{kl+l} \, \frac{k_2}{k_1} \cI_{\b(k+l) \dot{\r}(k+l)}  ({\bar{\bm{X}}}_1) \\
&&\quad \times \,\cI_{\ad(k)}{}^{\a(k)} (x_{3 \bar{1}}) \bar{\cI}^{\bd(k+l)\b(k+l)}(\tilde{x}_{\bar{3} 1}) \bar{\cI}_{\r(l)}{}^{\gd(l)}(\tilde{x}_{\bar{3} 1}) H_{\a(k) \bd(k+l),\, \gd(l)} (\fn3)~.~~ \non
\eea
On the other hand, from the second line of \eqref{inv-c3-2}, it follows that
\bea
\widetilde{H}_{\r(l)\dot{\r}(k+l), \,\ad(k)} (\bm{X}, \bm{\Q}, \bar{\bm \Q}) = H_{\r(l)\dot{\r}(k+l), \,\ad(k)} (-\bar{\bm{X}}, -\bm{\Q}, -\bar{\bm \Q})~,
\label{switch13-c3}
\eea
where \eqref{pt13} has been used. 

By virtue of eqs. \eqref{Htensor-c3b}, \eqref{b3-c3} and \eqref{b2-c3}, the right-hand side of \eqref{switch13-c3} takes the form
\bea
&&H_{(l, k+l , k)} (-\bar{\bm{X}}, -\bm{\Q}, -\bar{\bm \Q}; \, u, \bar{v}, \bar{w}) = u^{\r(l)} \bar{w}^{\dot{\r}(k+l)} \bar{v}^{\ad(k)}H_{\r(l)\dot{\r}(k+l), \,\ad(k)} (-\bar{\bm{X}}, -\bm{\Q}, -\bar{\bm \Q}) \non\\
&&= (-1)^l \bigg\{ a_1(\bar{v} \cdot \bar{w})^k \frac{\bar{\bm{X}}^l_{(1,1)}}{\bar{\bm X}^{2l+2}} - b_1  (\bar{v} \cdot \bar{w})^{k-1} (\bm{\Q} \bar{\bm{X}} \bar{w}) (\bar{v} \cdot \bar{\bm \Q}) \frac{\bar{\bm{X}}^l_{(1,1)}}{\bar{\bm{X}}^{2l+4}} \non\\
&&+ \frac{(\bar{v} \cdot \bar{w})^{k-1}}{k+l+1} (b_1 - 4\ri k a_1) \bigg[ (l+1)   (\bm{\Q} \bar{\bm{X}} \bar{v}) (\bar{w} \cdot \bar{\bm \Q}) \frac{\bar{\bm{X}}^l_{(1,1)}}{\bar{\bm{X}}^{2l+4}} \non\\
&&\qquad \qquad \qquad \qquad \qquad - l\,(\bar{v} \cdot \bar{w}) (u \cdot \bm{\Q}) (\bar{w} \cdot \bar{\bm \Q}) \frac{\bar{\bm{X}}^{l-1}_{(1,1)}}{\bar{\bm{X}}^{2l+2}} \bigg] \bigg\}~.
\label{517}
\eea
Direct computation of \eqref{tH-c3} gives rise to
\bea
&& \widetilde{H}_{(l, k+l , k)} (\fxq; \,u, \bar{v}, \bar{w}) =  u^{\r(l)} \bar{w}^{\dot{\r}(k+l)} \bar{v}^{\ad(k)} ~ \frac{(-1)^{kl+l}}{\bm{X}^2 \bar{\bm X}^{2l}} \non\\
&&\times \bigg\{ 
a_1 \ve_{\ad_1 \dot{\r}_1 } \dots \ve_{\ad_k \dot{\r}_k} \bar{\bm X}_{\r_1 \dot{\r}_{k+1}} \dots \bar{\bm X}_{\r_l \dot{\r}_{k+l}} \non\\
&&+\frac{b_1}{\bm{X}^2}\bm{\Q}^{\d} \bar{\bm \Q}^{\dot{\d}} \ve_{\ad_1 \dot{\r}_1 } \dots \ve_{\ad_k \dot{\r}_k} \bm{X}_{|\d| \dot{\r}_{k+1}} \bar{\bm X}_{\r_1 |\dot{\d}|} \bar{\bm X}_{\r_2 \dot{\r}_{k+2}} \dots \bar{\bm X}_{\r_l \dot{\r}_{k+l}} \non\\
&&-\frac{1}{{\bm X}^2} \frac{(4\ri l a_1 -b_1)}{k+l+1} \bigg[ (k+1) \ve_{\ad_1 \dot{\r}_1 } \dots \ve_{\ad_k \dot{\r}_k} \bm{X}^{\dot{\d} \d} \bm{\Q}_{|\d|} \bar{\bm \Q}_{\dot{\r}_{k+1}} \bar{\bm X}_{\r_1 \dot{\r}_{k+2}} \dots \bar{\bm X}_{\r_{l-1} \dot{\r}_{k+l}} \bar{\bm X}_{\r_l |\dot{\d}|}\non\\
&&\qquad \qquad- k \bar{\bm X}_{\r_1 \dot{\r}_1} \dots \bar{\bm X}_{\r_l \dot{\r}_l} \bm{\Q}^{\d} \bar{\bm \Q}^{\dot{\d}}  \bm{X}_{|\d| \ad_1} \ve_{|\dot{\d}| \dot{\r}_{l+1}} \ve_{\ad_2 \dot{\r}_{l+2} } \dots \ve_{\ad_k \dot{\r}_{k+l}} \bigg]
\bigg\}~.~~~
\eea
In order to compare this result with eq.~\eqref{517} we will use eq.~\eqref{XbarX} to express $\bm{X}$ in terms of $\bar{\bm X}$.
This gives
\bea
&&\widetilde{H}_{(l, k+l , k)} (\fxq; \, u, \bar{v}, \bar{w}) 
\non\\
&&= (-1)^{kl+l} \bigg\{ a_1(\bar{v} \cdot \bar{w})^k \frac{\bar{\bm{X}}^l_{(1,1)}}{\bar{\bm X}^{2l+2}} + (4\ri a_1+ b_1 ) (\bar{v} \cdot \bar{w})^{k-1} (\bm{\Q} \bar{\bm{X}} \bar{w}) (\bar{v} \cdot \bar{\bm \Q}) \frac{\bar{\bm{X}}^l_{(1,1)}}{\bar{\bm{X}}^{2l+4}} \non\\
&&- \frac{(\bar{v} \cdot \bar{w})^{k-1}}{k+l+1} \big(b_1 + 4\ri (k+1) a_1 \big) \bigg[ (l+1)   (\bm{\Q} \bar{\bm{X}} \bar{v}) (\bar{w} \cdot \bar{\bm \Q}) \frac{\bar{\bm{X}}^l_{(1,1)}}{\bar{\bm{X}}^{2l+4}} \non\\
&&\qquad \qquad \qquad \qquad \qquad \qquad \quad - l\,(\bar{v} \cdot \bar{w}) (u \cdot \bm{\Q}) (\bar{w} \cdot \bar{\bm \Q}) \frac{\bar{\bm{X}}^{l-1}_{(1,1)}}{\bar{\bm{X}}^{2l+2}} \bigg] \bigg\}~.
\label{519}
\eea
Equating \eqref{517} and \eqref{519}, it can be seen that the coefficients are constrained by
\bea
&&a_1 = (-1)^{kl}a_1~, \non\\
&&b_1 = (-1)^{kl+1} (4\ri a_1 +b_1)~, \qquad   b_1 - 4\ri k a_1 = (-1)^{kl+1}(b_1 + 4 \ri (k+1) a_1)~.~~~~
\label{520}
\eea
As a result, we have two scenarios to consider. 
The first case is when both $k$ and $l$ are odd. Here $a_1 = 0$ and $b_1$ is  an arbitrary complex parameter. 
The other case is when at least one of them is even, for which both parameters are now non-vanishing but related by $a_1 = \dfrac{\ri b_1}{2}$.

The three-point correlator of the higher-spin spinor current multiplets is thus determined up to a single complex coefficient $b_1 \equiv B$. It has the following structure
\bsubeq
\bea
\c3
&=& \frac{(x_{3 \bar{1}})_{(\ad_1}{}^{\a_1} \dots (x_{3 \bar{1}})_{\ad_k)}{}^{\a_k} (x_{2 \bar{3}})_{(\b_1}{}^{\bd_1} \dots (x_{2 \bar{3}})_{\b_{k+l})}{}^{\bd_{k+l}} }{(x_{\bar{1} 3}{}^2)^{k+1} x_{\bar{3}1}{}^2 x_{\bar{2} 3}{}^2(x_{\bar{3} 2}{}^2)^{k+l+1}}\non\\
&&\times H_{ \a(k) \bd(k+l), \, \gd(l)}
(\fn3) ~,
\eea
where the functional form of $H_{ \a(k) \bd(k+l), \, \gd(l)}$ depends on the values of $k$ and $l$. There are two different cases to consider:
\begin{itemize}
\item $\bm{(k,l)} =$ \textbf{(odd, odd)}
\bea
&&H_{(k,k+l,l)} (\fxq; \, u, \bar{v}, \bar{w})= B  (\bar{v} \cdot \bar{w})^{l-1} \bigg\{
(\bm{\Q} {\bm{X}} \bar{w}) (\bar{v} \cdot \bar{\bm \Q}) \frac{{\bm{X}}^k_{(1,1)}}{{\bm{X}}^{2k+4}} \non\\
&&+ \frac{(\bar{w} \cdot \bar{\bm \Q})}{k+l+1} \bigg[ (k+1)   (\bm{\Q} {\bm{X}} \bar{v})  \frac{{\bm{X}}^k_{(1,1)}}{{\bm{X}}^{2k+4}} - k\,(\bar{v} \cdot \bar{w}) (u \cdot \bm{\Q})  \frac{{\bm{X}}^{k-1}_{(1,1)}}{{\bm{X}}^{2k+2}} \bigg] \bigg\}~.~~~~~~~~
\eea
\item $\bm{(k,l)} =$ \textbf{(even, even), (odd, even), (even, odd)}
\bea
&&H_{(k,k+l,l)} (\fxq; \, u, \bar{v}, \bar{w})\non\\
&& = B  (\bar{v} \cdot \bar{w})^{l-1} \bigg\{ \frac{\ri}{2}(\bar{v} \cdot \bar{w})\frac{{\bm{X}}^k_{(1,1)}}{{\bm{X}}^{2k+2}} + (\bm{\Q} {\bm{X}} \bar{w}) (\bar{v} \cdot \bar{\bm \Q}) \frac{{\bm{X}}^k_{(1,1)}}{{\bm{X}}^{2k+4}} \non\\
&&+ \frac{2l+1}{k+l+1} (\bar{w} \cdot \bar{\bm \Q}) \bigg[ (k+1)   (\bm{\Q} {\bm{X}} \bar{v})  \frac{{\bm{X}}^k_{(1,1)}}{{\bm{X}}^{2k+4}} - k\,(\bar{v} \cdot \bar{w}) (u \cdot \bm{\Q})  \frac{{\bm{X}}^{k-1}_{(1,1)}}{{\bm{X}}^{2k+2}} \bigg] \bigg\}~.~~~~~~~~
\eea
\end{itemize}
\esubeq

In a more general setting, one may also consider
\bea
\la \bar{S}^{\prime}_{\ad(k)} (z_1) S_{\b(k+l)} (z_2) \bar{S}''_{\gd(l)}(z_3) \ra~,
\label{complex-c3}
\eea
where the correlator now does not satisfy \eqref{inv-c3} under the exchange $z_1 \leftrightarrow z_3$. As a result, one has to explicitly check the conservation equation at $z_3$, that is
\bea
\bar{D}_{(3)}^{\gd_1} \la \bar{S}^{\prime}_{\ad(k)} (z_1) S_{\b(k+l)} (z_2) \bar{S}''_{\gd(l)}(z_3) \ra = 0~, \quad D^2_{(3)} \la \bar{S}^{\prime}_{\ad(k)} (z_1) S_{\b(k+l)} (z_2) \bar{S}''_{\gd(l)}(z_3) \ra = 0~. ~~~~~
\label{524}
\eea
Making use of the relation \eqref{5.15} and the expression of $\widetilde{H}_{\r(l)\dot{\r}(k+l), \,\ad(k)} (\bm{X}_1, \bm{\Q}_1, \bar{\bm \Q}_1)$ which has been computed previously in \eqref{519}, it can be shown that \eqref{524} holds. Hence, the correlator \eqref{complex-c3} is  determined up to two \textit{complex} coefficients, $a_1$ and $b_1$.

\section*{Acknowledgements}
The work is supported in part by the Australian Research Council, project No. DP200101944. 


\begin{footnotesize}

\end{footnotesize}
\end{document}